\documentclass[a4paper,11pt]{article}
\pdfoutput=1 
\usepackage{jcappub} 
\usepackage[T1]{fontenc} 
\usepackage{graphicx}
\usepackage{array}
\usepackage{hyperref}
\usepackage{setspace}
\usepackage{mathtools}
\usepackage{tensor}
\usepackage{booktabs}
\usepackage[dvipsnames]{xcolor}
\usepackage{makecell}
\usepackage{amsmath, amsfonts}
\usepackage[utf8]{inputenc}
\usepackage[english]{babel}
\renewcommand{\sideset}[3]{\tensor[#1]{#3}{#2}}

\newcolumntype{M}[1]{>{\centering\arraybackslash}m{#1}}
\newcolumntype{N}{@{}m{0pt}@{}}

\title{Cosmological Forecast for non-Gaussian Statistics in large-scale weak Lensing Surveys}

\author[a,1]{Dominik Z\"urcher, \note{Corresponding Author, Email: dominik.zuercher@phys.ethz.ch }}
\author[a]{Janis Fluri,}
\author[a]{Raphael Sgier,}
\author[a]{Tomasz Kacprzak,}
\author[a]{Alexandre Refregier}
\affiliation[a]{Institute for Particle Physics and Astrophysics, Department of Physics, ETH Z\"urich, \\Wolfgang Pauli Strasse 27, 8093 Z\"urich, Switzerland}

\abstract{Cosmic shear data contains a large amount of cosmological information encapsulated
in the non-Gaussian features of the weak lensing mass maps.
Weak lensing studies mostly rely on two-point
statistics to constrain cosmology from cosmic shear data, that do not capture all of this information.
Additional non-Gaussian information can be extracted using non-Gaussian statistics.
We compare the constraining power in
the $\Omega_{\mathrm{m}} - \sigma_8$ plane of three map-based non-Gaussian statistics with the angular power spectrum,
namely; peak counts, minimum counts and Minkowski functionals.
We further analyze the impact of tomography
and systematic effects originating from galaxy intrinsic alignments, multiplicative
shear bias and photometric redshift systematics.
We forecast the performance of the statistics for a stage-3-like
weak lensing survey, spanning an area of 5000 deg$^2$ and restrict ourselves to
scales $\geq$ 10 arcmin to avoid baryonic effects.
The study follows a forward
modelling scheme to predict the statistics at different cosmologies based on N-Body simulations.
We find, that in our setup, the considered non-Gaussian statistics provide tighter constraints
than the angular power spectrum. The peak counts show the
greatest potential, increasing the Figure-of-Merit (FoM) in the $\Omega_{\mathrm{m}} - \sigma_8$ plane by
a factor of about 4, while the minimum counts and the Minkowski functionals yield
an increase by a factor of about 2. A combined analysis using all non-Gaussian statistics in addition
to the power spectrum increases the FoM by a factor of 5 and reduces the error on $S_8$ by $\approx$ 25\%.
We find that
the importance of tomography is diminished when combining non-Gaussian statistics
with the angular power spectrum. The non-Gaussian statistics indeed profit less from
tomography and the minimum counts and Minkowski functionals add some robustness
against galaxy intrinsic alignment in a non-tomographic setting. We further find
that a combination of the angular power spectrum and the
non-Gaussian statistics allows us to apply conservative scale cuts in the analysis,
thus helping to minimize the impact of baryonic and relativistic effects, while conserving the cosmological
constraining power. We make the code that was used to conduct this analysis publicly
available to simplify performing such analyses in the
future\footnote{\texttt{NGSF}: \url{https://cosmo-gitlab.phys.ethz.ch/cosmo_public/NGSF}}.}

\begin{document}
\maketitle
\flushbottom

\section{Introduction}
\label{sec:Introduction}
The $\Lambda$CDM model succeeds to explain
and predict the main observables of the Universe, including the Big Bang
nucleosynthesis (see e.g. \cite{fields2006big}), the anisotropies of the
cosmic microwave background (CMB) (see e.g. \cite{hinshaw2013nine}) and the Hubble
diagram of Type Ia supernovae (see e.g. \cite{betoule2014improved}).
Although recent results point towards a disagreement in the value of the Hubble parameter
as inferred from local measurements \cite{riess20162, riess2018milky,
bonvin2017h0licow, birrer2019h0licow}
and from CMB studies \cite{ade2016planck, aghanim2018planck},
the $\Lambda$CDM model remains the most successful cosmological model to date.
It is based on general relativity and mainly characterized by a flat geometry,
a cosmological constant $\Lambda$ and a cold dark matter (CDM) component, driving
the formation of structures. To further deepen our
understanding of the Universe, it is essential to provide novel measurements
that are able to challenge the model and potentially discover deviations from
it, that might lead to the discovery of new physics. \\

\noindent One way to provide such tests is given by the investigation of the cosmic shear, which is the
coherent distortion of the apparent ellipticities of galaxies, caused by weak gravitational lensing (WL) by
the foreground large-scale structure of the Universe \cite{kilbinger2015cosmology}.
These distortions are typically at the percent level.
However, by measuring millions of galaxy shapes on the sky, as it
is achieved by modern large scale imaging surveys, the statistical potential of
the method is large. The simple theoretical description of
WL, as well as its independence on galaxy biasing,
are further advantages of the method \cite{force2006report}. \\

\noindent The feasibility and potential of cosmic shear studies
was successfully demonstrated by past surveys, such as the
Canada France Hawaii Telescope Lensing Survey (CFHTLens) \cite{heymans2013cfhtlens}
or the Sloan Digital Sky Survey (SDSS) \cite{huff2012cosmic}. Putting new constraints
on the cosmological model
using WL is not only one of the main science goals of currently ongoing stage 3 surveys, such as
the Dark Energy Survey (DES) \cite{abbott2018dark}, Kilo-Degree Survey (KIDS) \cite{hildebrandt2016kids}
or the Hyper Suprime-Cam (HSC) \cite{hikage2019cosmology},
but also served as one of the major motivations for future stage 4 surveys such as the
Large Synoptic Survey Telescope (LSST) \cite{abell2009lsst} or Euclid \cite{amendola2018cosmology}. \\

\noindent Cosmic shear measurements are affected by a variety of systematic effects.
The accessibility of small scales in cosmic shear experiments is limited due to biases
arising from baryonic physics, such as radiative cooling (see e.g. \cite{yang2013baryon})
or feedback effects caused by the active galactic nucleus (AGN), stellar winds or supernovae \cite{osato2015impact}.
These baryonic effects are generally
difficult to treat in a dark-matter-only framework, as it is commonly used in cosmic shear studies
(see e.g. \cite{osato2015impact}). Galaxy intrinsic alignment, describing the gravitational
interaction of galaxies with the large-scale structure, can lead to correlations
of the intrinsic ellipticities of the source galaxies
and a contamination of the cosmic shear signal (see e.g. \cite{heavens2000intrinsic}).
This effect is similarly difficult to account for as baryonic effects.
In addition to systematic effects arising from unaccounted physics, biases can
arise due to imperfections in the measurement and data reduction process.
Some of these effects can be taken into account by introducing a multiplicative shear bias
(see e.g. \cite{voigt2010limitations}). In particular, inaccuracies in the measurement of the redshift
distribution of the source galaxies can bias cosmic shear
measurements. These photometric redshift errors cannot be modelled as
a multiplicative bias component and are therefore treated separately
(see e.g. \cite{bonnett2015redshift}).
Further higher-order systematic effects include magnification bias,
source-lens clustering or reduced shear bias, for example.\\

\noindent As the number of measured galaxy shapes increases with the observed
cosmological volume, the statistical error in the measurements decreases
and the correct understanding and
treatment of these systematic effects becomes more pressing.
Also, with the advent of tensions between the inferred values of cosmological
parameters from different cosmological probes, finding new ways to improve the
robustness of WL measurements against these systematic effects becomes
essential. An important example is the disagreement in the value of the
amplitude of density fluctuations $\sigma_8$,
between WL surveys and the results from CMB experiments like Planck \cite{aghanim2018planck},
with the WL studies yielding consistently lower values \cite{maccrann2015cosmic}. \\

\noindent Up to now, the shear two-point correlation function and its Fourier counterpart, the
angular power spectrum, served as the main WL observables. While ongoing surveys
like KIDS, DES and HSC make major contributions to the understanding of how systematics
affect those two-point statistics, we are
taking a complementary route by investigating the robustness of alternative, non-Gaussian
WL observables to the major effects driving the systematic uncertainty. \\

\noindent In the case of a homogeneous, isotropic Gaussian random field two-point
statistics are sufficient summary statistics.
However, due to the non-linear nature of gravitational collapse, this assumption
is not valid on small scales and at late times, as
the density field becomes non-Gaussian.
Therefore, two-point statistics are insufficient to fully describe the matter
density field and additional statistics
ought to be considered (see e.g. \cite{petri2014impact}).
Additionally, each statistic is affected differently by systematic effects.
Hence, a combination of multiple statistics can allow for a better calibration
and understanding of the different systematics and ultimately improve the
robustness of the measurement. \\

\noindent A variety of statistics optimized to capture the
non-Gaussian information of the cosmic shear signal was previously developed and tested.
A natural extension after the study of two-point statistics
is to consider higher-order statistics, such as three-point correlation functions \cite{semboloni2011weak}
or the bispectrum \cite{cooray2001weak},
which has been shown to capture complementary information and significantly improve
parameter constraints \cite{takada2004cosmological}. A computationally less demanding way to access
the additional information of the cosmic shear signal is provided by the moments
of WL mass maps. This method was first studied by \cite{jain1997cosmological}
and a recent study demonstrates its potential for the Year 3 data of the
DES \cite{gatti2019dark}. Other non-Gaussian statistics include for example,
higher moments (see e.g. \cite{chang2018dark})
or the PDF of the convergence field \cite{patton2017cosmological}.
In this work, we forecast the performance of peak counts (PC)
(see Section \ref{sec:paf}), minimum counts (MC) (see Section \ref{sec:vaf})
and Minkowski Functionals (MFs) (see Section \ref{sec:mf}). \\

\noindent It was shown, by \cite{kratochvil2010probing} that the count of peaks on mass maps
can be used to constrain cosmology. The method was applied to CFHTLens data by
\cite{liu2015cosmology} and to the DES Science Verification data by \cite{kacprzak2016cosmology}.
The potential of using the lensing signal around local minima of WL mass maps
was demonstrated by \cite{sanchez2016cosmic} and \cite{gruen2016weak},
using the DES Science Verification data. The counts of such local minima of the mass maps
can also serve as a mean to constrain cosmology \cite{coulton2019weak}.
While the PC and MC focus on extracting information from local features of the lensing
signal, it was shown, that MFs can probe cosmology
using the global topology of mass maps \cite{kratochvil2012probing}.
The method was applied
to CFHTLens data by \cite{petri2015emulating}. \\

\noindent The primary goal of this work is to forecast and compare the resulting
constraining power on the total matter density $\Omega_{\mathrm{m}}$
and the fluctuation amplitude $\sigma_8$ when using
the PC, MC, MFs, as well as the angular power spectrum and different
combinations of these statistics. We forecast
the performance, using a simulated stage-3-like WL survey. \\

\noindent Additionally, we investigate the robustness of the statistics against the three
major WL systematics; galaxy intrinsic alignment, multiplicative
shear bias and photometric redshift error (see Section \ref{sec:systematics}).
The two-point statistics are affected by these effects.
In particular, galaxy intrinsic alignment has been shown
to potentially bias the cosmological constraints and increase systematic
uncertainty, with the two-point statistics being unable
to constrain galaxy intrinsic alignment in a non-tomographic setup (see e.g. \cite{bridle2007dark}).
While the use of tomography improves the constraints on galaxy intrinsic alignment
considerably and helps to reclaim a large part of the otherwise lost constraining power,
it requires knowledge
of the redshift distribution of the source galaxy population, obtained through photometric
redshift estimates \cite{heymans2013cfhtlens}.
With photometric redshift estimation being a source of
systematic biases and uncertainty itself, we investigate the possibility of non-Gaussian
statistics being able to constrain galaxy intrinsic alignment in a non-tomographic analysis. \\

\noindent While a lot of information about the large-scale structure can be
obtained from very large- and small-scale features of the cosmic shear
field, additional systematic effects become relevant on these scales
(e.g. relativistic corrections on large scales \cite{giblin2017general} or
baryonic corrections on small scales \cite{osato2015impact}).
This limits the range of scales that can be safely considered in cosmic shear analyses.
We study, whether the addition of non-Gaussian statistics to the analysis can provide
an alternative route to using such scales by providing additional constraining power. \\

\noindent Contrarily to the two-point statistics, non-Gaussian statistics have
the disadvantage that making analytical predictions from theory for different
cosmologies is often difficult, complicating the inference of cosmological constraints.
One approach is to rely on analytical approximations to predict the statistics
at different cosmologies, as it is done for peak counts in \cite{maturi2010analytic}
or for Minkowski functionals in \cite{parroni2020going}, for example.
We take a different route and circumvent this problem in our analysis
by relying on a forward modelling approach,
predicting the statistics for different cosmologies
based on a suite of N-Body simulations and therefore avoiding the necessity of
analytical predictions of the statistics.
However, this approach has the disadvantage that it comes with a
higher computational cost and requires to setup a simulation pipeline.
Hence, we develop and distribute a code framework aimed at simplifying
this kind of WL analysis. \\

\noindent We start with a summary of the most important properties of
WL and an introduction of the studied statistics in Section \ref{sec:theory}.
In Section \ref{sec:methodology}, we guide the reader through our analysis and introduce
the ingredients and codes used in this work. We present the simulated, non-tomographic
statistics and their cosmology-dependence in Section \ref{sec:statistics} and follow
up with an investigation on their cosmological constraining power and robustness
to the studied systematics in Section \ref{sec:constrains}.
The work concludes with the main findings and a short outlook on possible extensions
in Section \ref{sec:conclusions}.

\section{Weak Lensing Statistics}
\label{sec:theory}
The phenomenon of gravitational lensing refers to the deflection of
photons, traveling from a distant source towards an observer. The deflection is
caused by the foreground density fluctuations along the line of sight.
In the context of gravitational lensing
the foreground density fluctuations act like a medium with variable refractive
index for the propagating photons, causing their deflection.
Gravitational lensing can cause the appearance of an extended background object to be distorted.
In the regime of WL, where the distortions are small,
the change of the shape of a lensed object can be broken down into two parts; the convergence $\kappa$,
describing an isotropic magnification of the object and the shear $\gamma = (\gamma_1, \gamma_2)$
denoting an anisotropic stretching (see e.g. \cite{bartelmann2001weak, kilbinger2015cosmology}). \\

\noindent In cosmic shear studies one observes the percent level distortions of
the ellipticities of distant galaxies, caused by the lensing due to the foreground
large-scale structure of the Universe.
Therefore, by measuring the fields $\gamma$ and $\kappa$ on the sky,
which are related to the gravitational potential $\Phi$ that is
induced by the large-scale structure, one can learn about the distribution of
matter in the local Universe. Due to this connection,
cosmic shear is mostly sensitive to the cosmological parameters describing
the matter distribution of the local Universe, namely; the matter density $\Omega_{\mathrm{m}}$,
the fluctuation amplitude $\sigma_8$,
the dark energy density $\Omega_{\Lambda}$ and the dark energy equation-of-state parameter $w$
(see e.g. \cite{huterer2006systematic}). \\

\noindent The two fields $\kappa$ and $\gamma$ are not independent of each other,
but they are linked via the gravitational potential $\Phi$ \cite{wallis2017mapping}.
This connection can be used to derive $\kappa$ from $\gamma$ and vice versa.
The most widely used method to do so
is the Kaiser-Squires mass-mapping method \cite{kaiser1993mapping}.
This approach relies on approximating part of the celestial sphere as a plane.
While this assumption is valid for small scale surveys like CFHTLens,
it is not applicable for ongoing stage 3 surveys like DES \cite{abbott2018dark}
or stage 4 surveys like LSST \cite{abell2009lsst}. Therefore, we
rely on a spherical extension of this method introduced in \cite{wallis2017mapping}. \\

\noindent Being defined as fields on the sphere, $\kappa$ and $\gamma$ can be
decomposed in the basis of spherical harmonics as
\begin{align}
\kappa(\theta, \phi) &= \sum_{\ell=0}^{\infty} \sum_{m=-\ell}^{\ell} \sideset{_0}{_{\ell m}}{\hat{\kappa}} \cdot \sideset{_0}{_{\ell m}}Y(\theta, \phi),
\label{eq:decomp} \\
\gamma(\theta, \phi) &= \sum_{\ell=0}^{\infty} \sum_{m=-\ell}^{\ell} \sideset{_2}{_{\ell m}}{\hat{\gamma}} \cdot \sideset{_2}{_{\ell m}}Y(\theta, \phi),
\end{align}
where $\sideset{_0}{_{\ell m}}Y(\theta, \phi)$ are the spin-0 and $\sideset{_2}{_{\ell m}}Y(\theta, \phi)$ the spin-2 spherical
harmonics, respectively. The coefficients $\sideset{_0}{_{\ell m}}{\hat{\kappa}}$ can be calculated from
the $_2\hat{\gamma}_{\ell m}$ coefficients via the relation
\begin{equation}
_0\hat{\kappa}_{\ell m} = D_{\ell}^{-1} \cdot \sideset{_2}{_{\ell m}}{\hat{\gamma}},
\label{eq:KS}
\end{equation}
using the kernel
\begin{equation}
D_{\ell} = \frac{-1}{\ell(\ell + 1)}\sqrt{\frac{(\ell + 2)!}{(\ell - 2)!}}.
\end{equation}
This relation can be obtained by relating the coefficients via the gravitational
potential $\Phi$ \cite{wallis2017mapping}.
The inverse relation can also be used to reconstruct the $\gamma$ field from
the $\kappa$ field in the same fashion.
This approach was already used successfully in \cite{fluri2018weak} and \cite{chang2018dark}, for example. \\

\subsection{Angular Power Spectrum (CLs)}
\label{sec:cls}
The angular two-point correlation function and its Fourier analogue, the
angular power spectrum (CLs), served as the main statistics for the extraction of
information from cosmic shear data in past WL surveys
(see e.g. \cite{hildebrandt2016kids, heymans2013cfhtlens}).
The angular two-point correlation function
\begin{equation}
\xi(|\vec{x}_1 - \vec{x}_2|=\theta) = \left\langle F(\vec{x}_1), F(\vec{x}_2)\right\rangle
\end{equation}
describes the expected value of a random field $F$ at a fixed angular distance $\theta$
from a random point $\vec{x}_1$, given that a certain value of the field was
measured at that point \cite{peebles1980large}.
Given a decomposition of the field $F$ in the basis of spherical harmonics as
\begin{equation}
F = \sum_{\ell=0}^{\infty} \sum_{m=-\ell}^{\ell} a_{lm} \sideset{}{_{\ell m}}Y(\theta, \phi),
\end{equation}
one can define the angular power spectrum $C_{\ell}$ of the field $F$ as
\begin{equation}
C_{\ell} = \frac{1}{2\ell + 1}\sum_{m=-\ell}^{\ell} |a_{lm}|^2.
\end{equation}
The convergence field $\kappa$ is commonly decomposed into a curl-free
component $\kappa_{\mathrm{E}}$ and a divergence-free component $\kappa_{\mathrm{B}}$ as
\begin{equation}
\kappa(\theta, \phi) = \kappa_{\mathrm{E}} + i\kappa_{\mathrm{B}} = \sum_{\ell=0}^{\infty} \sum_{m=-\ell}^{\ell} (\sideset{_0}{_{\mathrm{E}, \ell m}}{\hat{\kappa}} + i\sideset{_0}{_{\mathrm{B}, \ell m}}{\hat{\kappa}} )  \enspace \sideset{_0}{_{\ell m}}Y(\theta, \phi),
\end{equation}
allowing the decomposition of the angular power spectrum into three separate terms
\begin{align}
C_{\ell}^{\mathrm{EB}} &= \frac{1}{2\ell + 1}\sum_{m=-\ell}^{\ell}  \sideset{_0}{_{\mathrm{E}, \ell m}}{\hat{\kappa}}\sideset{_0}{_{\mathrm{B}, \ell m}^*}{\hat{\kappa}}, \\
C_{\ell}^{\mathrm{EE}} &= \frac{1}{2\ell + 1}\sum_{m=-\ell}^{\ell}  |\sideset{_0}{_{\mathrm{E}, \ell m}}{\hat{\kappa}}|^2, \\
C_{\ell}^{\mathrm{BB}} &= \frac{1}{2\ell + 1}\sum_{m=-\ell}^{\ell}  |\sideset{_0}{_{\mathrm{B}, \ell m}}{\hat{\kappa}}|^2. \\
\end{align}
The vast majority of the cosmological signal is carried by the E-modes ($C_{\ell}^{EE}$),
while B-modes ($C_{\ell}^{\mathrm{BB}}$) mostly arise from systematics in the shear-calibration
process or object selection biases. EB-modes ($C_{\ell}^{\mathrm{EB}}$) are generated
via mode mixing due to masking effects.
Given that most of the cosmological signal is captured
by the E-modes, we neglect EB-, and B-modes in this study.
The angular power spectrum can also be related to an integral over the matter power spectrum
of the Universe \cite{nicola2014three}.
Using the Limber and small-angle approximations, the computation can be sped up
significantly and allows to obtain approximative predictions for the CLs for different
cosmologies (see e.g. \cite{bartelmann2001weak, limber1953analysis}).
We note that we do not require to denoise the measurement CL signal in this work, as it is commonly
done in analyses where the CL signal needs to be compared to a theory prediction.
In a simulation-based approach, as it is used in this work, the measurement
and the predictions of the signal at different cosmologies both contain a statistically equivalent
noise component (see Section \ref{sec:methodology}).

\subsection{Peak Counts (PC)}
\label{sec:paf}
The idea that massive dark matter halos could imprint themselves onto mass maps
as local maxima, so called peaks, was pioneered by the works of
\cite{kaiser1993mapping, tyson1990detection, miralda1991correlation}. While peaks were first studied mainly
as a mean to detect massive clusters from mass maps (see e.g. \cite{schneider1996detection}),
it was found later that they can also serve directly as a cosmological probe (see e.g. \cite{reblinsky1999cosmic, liu2016origin}).
We detect peaks from the pixelized mass maps by comparing each pixel to its direct neighbors.
A pixel is regarded as a peak if its value is higher than all of the
values of its neighboring pixels.
We bin the found peaks as a function of their convergence value.
In addition to counting peaks, the consideration of further peak statistics, such as
peak-profiles or peak-correlation functions can help to improve cosmological constraints
\cite{marian2013cosmological}. In this work we only study peak counts (PC) and we leave the
exploration of peak-profiles and peak-correlation functions to further studies.
While using peaks instead of CLs has the advantage of becoming more sensitive to
the non-Gaussian features of the maps \cite{berge2010optimal},
it comes at the cost of a complicated and at most approximative analytical prediction
of the PC for different cosmologies (see e.g. \cite{maturi2010analytic}).
We avoid having to rely on such approximative predictions by using a forward
modelling approach and predict the PC for different cosmologies
using simulations, as described in Section \ref{sec:methodology}.

\subsection{Minimum Counts (MC)}
\label{sec:vaf}
While the idea of using peaks as a cosmological probe became popular in
recent years, using counts of local minima of the mass
maps to infer cosmology received less attention, although the lensing signal around such under-dense regions
was already proposed as a way to provide insight into interesting physics,
such as modified gravity (see e.g. \cite{baker2018void, paillas2019santiago}).
While peaks are sensitive to over-densities of the matter distribution,
local minima probe its under-densities.
Hence, they can potentially probe complementary information.
Another aspect that makes local minima an interesting probe, is that they target
regions with small amounts of baryonic matter. It was
shown, that local minima suffer less from effects related to baryonic physics
than other statistics \cite{coulton2019weak}.
The identification of local minima of the projected WL signal, as compared to
finding under-dense regions from the three dimensional matter distribution, has the
advantage that one does not require a complicated void identification scheme nor
a void tracer, such as halos \cite{coulton2019weak}.
Our detection of local minima is similar
to the detection of peaks. We record a pixel as a minimum, if the recorded value is
smaller than the values recorded for all of its direct neighbors.
The same kinds of summary statistics as for peaks can be used for minima as well;
minimum counts (MC), the profiles around minima and the correlation function of minima.
We only study the MC and leave the investigation of the other statistics to future studies.

\subsection{Minkowki Functionals (MFs)}
\label{sec:mf}
Minkowski Functionals (MFs) are mathematical descriptors of the global topology of continuous,
stochastic fields. They capture information contained in the n-point correlators of the field of any order n,
which makes them natural probes of non-Gaussianity \cite{mecke288buchert}.
For a two-dimensional field, such as a mass map, there exist only three MFs, dubbed $V_0$, $V_1$ and $V_2$.
As the MFs are scale-dependent, they are calculated for a number of different excursion sets $Q_{t}$ of the field.
The excursion set $Q_{t}$ is formed by the region of the field where the field value
exceeds the threshold $t$. Hence, the MFs are functions of $t$.
The three MFs for a two-dimensional field are defined as
\begin{align}
V_0(t) &= \int_{Q_{t}} \mathrm{d}a ,\\
V_1(t) &= \frac{1}{4}\int_{\partial Q_{t}} \mathrm{d}l ,\\
V_2(t) &= \frac{1}{2\pi}\int_{\partial Q_{t}} \mathcal{K} \thinspace \mathrm{d}l,
\end{align}
where $\mathrm{d}a$ and $\mathrm{d}l$ are the surface and line elements of the
excursion sets $Q_{t}$ and $\mathcal{K}$ is the local geodesic curvature.
Geometrically speaking, $V_0$ describes the area, $V_1$ the perimeter and $V_2$
the Euler characteristic of the excursion sets \cite{vicinanza2019minkowski}.
Since the MFs can be analytically computed for a Gaussian random field,
they are commonly used to quantify the deviation from Gaussianity of a field
\cite{tomita1986curvature}.
On the other hand, no exact, analytical prediction can be made for non-Gaussian fields.
Commonly, the non-Gaussian part of the MFs is treated as a perturbation of the Gaussian
part and expanded in a perturbation series
(see e.g. \cite{vicinanza2019minkowski, parroni2020going, petri2013cosmology}).
If the non-Gaussianity of the field is weak, the series converges and can be truncated
to obtain an approximative prediction for the MFs. In the presence of strong non-Gaussianity
though, as it is typically the case for mass maps, the series does not converge \cite{petri2013cosmology}.
In our forward modelling approach, we are not affected by this problem, since we do not
require to make analytical predictions for the MFs at different cosmologies.
Following \cite{petri2013cosmology}, we measure the MFs from the mass maps directly as
\begin{align}
V_0(t) &= \frac{1}{Q_{t}}\int_{Q_{t}} \Theta(\kappa(\vec{x}) - t) \thinspace \mathrm{d}x \thinspace \mathrm{d}y ,\\
V_1(t) &= \frac{1}{4Q_{t}}\int_{Q_{t}} \delta(\kappa(\vec{x}) - t)\sqrt{(\partial_x \kappa)^2 + (\partial_y \kappa)^2} \thinspace \mathrm{d}x \thinspace \mathrm{d}y ,\\
V_2(t) &= \frac{1}{2\pi Q_{t}}\int_{Q_{t}} \delta(\kappa(\vec{x}) - t) \frac{2\partial_x \kappa \partial_y \kappa \partial_x \partial_y \kappa - (\partial_x \kappa)^2 \partial_y^2 \kappa - (\partial_y \kappa)^2 \partial_x^2 \kappa}{(\partial_x \kappa)^2 + (\partial_y \kappa)^2} \thinspace \mathrm{d}x \thinspace \mathrm{d}y,
\end{align}
where $\Theta$ and $\delta$ denote the Heaviside step function and the Dirac
delta function, respectively. The gradients are calculated numerically on a pixel level.

\section{Method}
\label{sec:methodology}
In this work, we are forecasting the constraining power of different map-based
statistics for a stage-3-like WL survey and we investigate their robustness to systematic effects.
In the forward modelling framework that we have chosen to conduct this study we
require a stage-3-like mock survey and a suite of N-Body simulations
spanning a range of different cosmologies. Using these two ingredients we simulate
mass maps with the same survey properties as the mock survey but with different
cosmological signals, by drawing the noise signal from the mock survey and adding
it to the simulated cosmological signal. These maps allow us to predict the statistics
and calculate the likelihood at different cosmologies enabling us to use Bayesian
inference to find the parameter constraints. \\
In the following we describe the different steps involved in this process in greater detail.
Our analysis is built on the work of \cite{fluri2018weak}, where a similar approach
was used to investigate the constraining power of the peak abundance function
for a 2000 deg$^2$ survey.

\subsection{Mock Survey}
We generate a stage-3-like mock survey by randomly drawing galaxy positions on the sky.
The positions are drawn within a square patch (in a cylindrical projection) of 5000 deg$^2$
until the target galaxy density of 5 arcmin$^{-2}$ is reached.
For each galaxy, we randomly draw its ellipticity $e$ from a probability density given as
\begin{equation}
\mathrm{Prob}(|e|) \propto (|e| + 0.01)^{-4}[1 - \exp(-23 |e|^4)].
\end{equation}
This function was chosen to fit the distribution of the galaxy ellipticities recorded
in \cite{troxel2018dark} (hereafter T18). The functional form was
proposed by \cite{bruderer2016calibrated} and used successfully in \cite{fluri2018weak}.
The individual ellipticity components $e_1$ and $e_2$ are obtained by random rotation
of the ellipticity as
\begin{align}
e_1 &= \Re{[|e|\exp(i\phi)]}, \nonumber\\
e_2 &= \Im{[|e|\exp(i\phi)]}, \label{eq:rotation}
\end{align}
where the angle $\phi$ is drawn uniformly from the interval $[0, 2\pi[$.
We truncate the ellipticity distribution at a value of 1.5 in order to avoid extreme outlier events.
Note that the mock survey does not need to contain a cosmological signal, since it is only used as a source
for the noise component (see Section \ref{sec:mass_maps}). \\

\begin{figure}
\centering
\includegraphics[width=0.8 \textwidth]{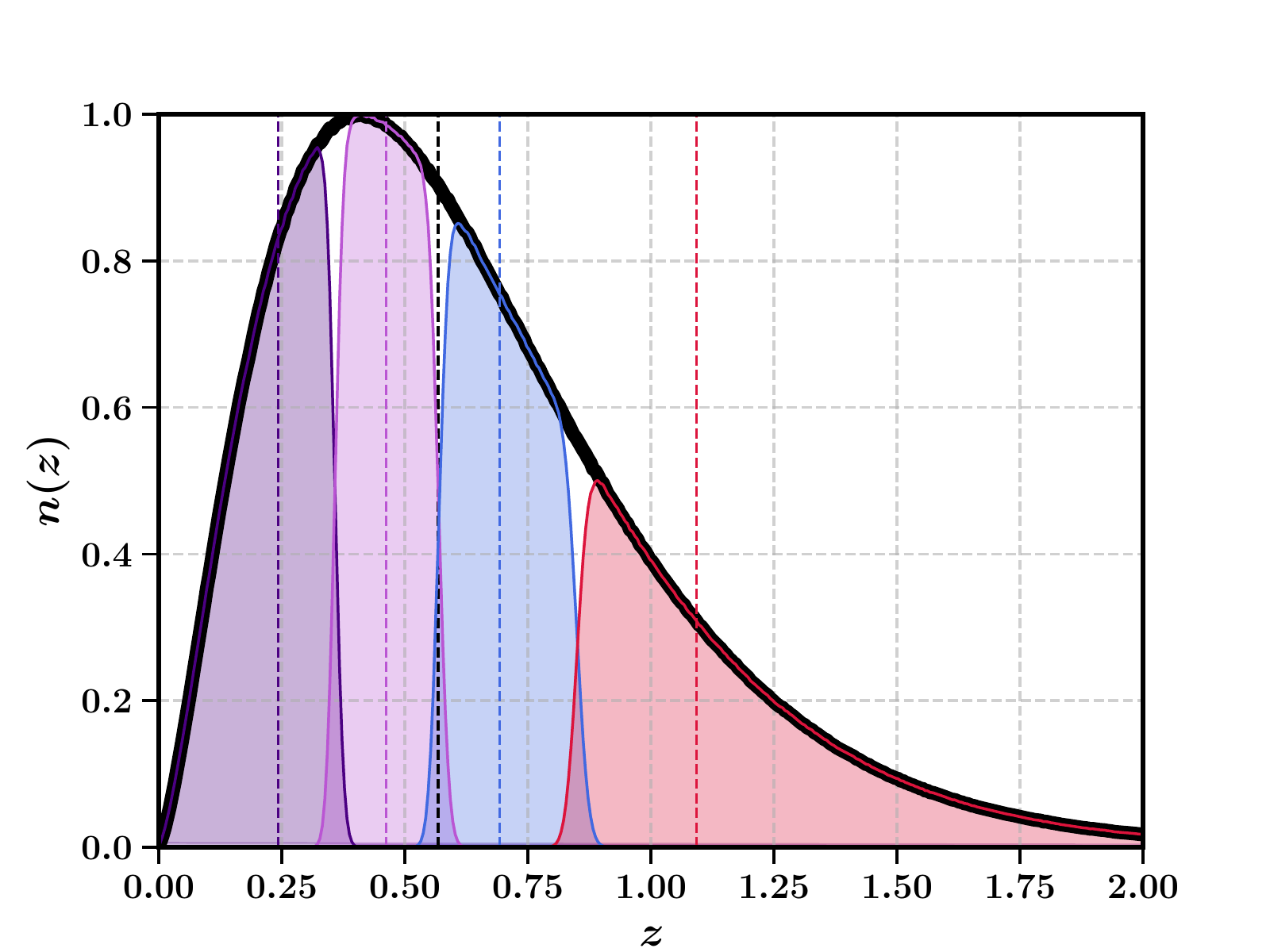}
\caption{\label{fig:nz} Global redshift distribution $n(z)$ of the simulated galaxies in the
mock survey as well as the distributions of the 4 tomographic bins.
The global redshift distribution as indicated by the black curve was fitted to the
redshift distribution found by T18. The dashed, vertical lines indicate the median
redshift of the global and tomographic distributions, respectively. The distributions are normalized.}
\end{figure}

\noindent The redshift $z$ of the galaxies is drawn from a Smail distribution \cite{smail1995deep} parameterized as
\begin{equation}
\mathrm{Prob}(z) \propto z^{1.5}\exp\left(-\left[\frac{z}{0.31}\right]^{1.1}\right),
\end{equation}
where we fitted the parameters, such that the global redshift distribution $n(z)$
of the simulated galaxies resembles the redshift distribution found in T18.
Since we also aim to compare parameter constraints in a tomographic setup, we
additionally require to assign each galaxy to a tomographic bin.
We perform this division, such that each tomographic bin contains the same number of galaxies,
according to the scheme used in \cite{amara2007optimal}.
Following the analysis of T18 we use 4 tomographic bins for the tomographic setup.
The global redshift distribution $n(z)$ of the simulated galaxies, as well as the
distributions of the 4 tomographic bins is shown in Figure \ref{fig:nz}.

\subsection{N-Body Simulations}
\label{sec:simulations}
We use the publicly available N-Body code \texttt{PKDGRAV3}
\cite{potter2017pkdgrav3} to sample a grid of cosmologies in the $\Omega_{\mathrm{m}} - \sigma_8$ plane.
\texttt{PKDGRAV3} is a dark-matter-only, full-tree code that uses a fast multipole expansion to
calculate the gravitational force, achieving a linearly increasing run time
in the number of particles. The code also features GPU-acceleration. \\

\noindent The simulations used in this work were performed using $768^3$ particles,
a box with a side-length of 900 Mpc/h and periodic boundary conditions.
Depending on cosmology, we replicate the box up to 14 times along each dimension ($14^3$ replications in total),
in order to sample a large enough cosmological volume, such that we can cover the
necessary redshift range (up to $z=3.0$). Note that such a replication under-predicts
the variance of the simulations on large scales. However, since we use a
lower scale cut of $\ell=100$ the results are not affected by the replication
(see Section \ref{sec:cls}).
The initial conditions for the simulations are generated at $z=99.0$.
The resulting particle positions are returned in 87 shells taken
from redshift $z=3.0$ up to redshift $z=0.0$ using the lightcone mode of \texttt{PKDGRAV3}.
We note, that due to the inner workings of \texttt{PKDGRAV3}, the shells are
not spaced equally in redshift and their location is also slightly varying with cosmology.
The default precision settings of \texttt{PKDGRAV3} were used. \\

\noindent We adopt a flat $\Lambda$CDM cosmology and we fix
all cosmological parameters except for $\Omega_{\mathrm{m}}$ and $\sigma_8$ to the
($\Lambda$CDM,TT,TE,EE+lowE+lensing) results from Planck 2018 \cite{aghanim2018planck} for all simulations.
This corresponds to a dimensionless Hubble parameter $h=0.6736$,
dark energy equation-of-state parameter $w=-1$, baryon density $\Omega_{\mathrm{b}}=0.0493$
and a scalar spectral index $n_{\mathrm{s}}=0.9649$. The dark energy density $\Omega_{\Lambda}$
was chosen depending on the value of $\Omega_{\mathrm{m}}$ such that a flat geometry is realized.
We include massive neutrinos in our simulations, adapting
a degenerate mass hierarchy with a minimal neutrino mass of $m_{\nu}=0.02$ eV
per neutrino in all simulations. The neutrinos are treated as a relativistic fluid,
according to the scheme outlined in \cite{tram2019fully}.
This results in a neutrino energy density of $\Omega_{\nu} \approx 0.0014$ at present time.
We note, that we have subtracted $\Omega_{\nu}$ from the initial dark matter energy density $\Omega_{\mathrm{CDM}}$.
Therefore, all the values of $\Omega_{\mathrm{m}}$ reported in this work should be interpreted
as a sum of the three contributions from $\Omega_{\mathrm{CDM}}$, $\Omega_{\nu}$ and $\Omega_{\mathrm{b}}$. \\

\begin{figure}
\centering
\includegraphics[width=0.8 \textwidth]{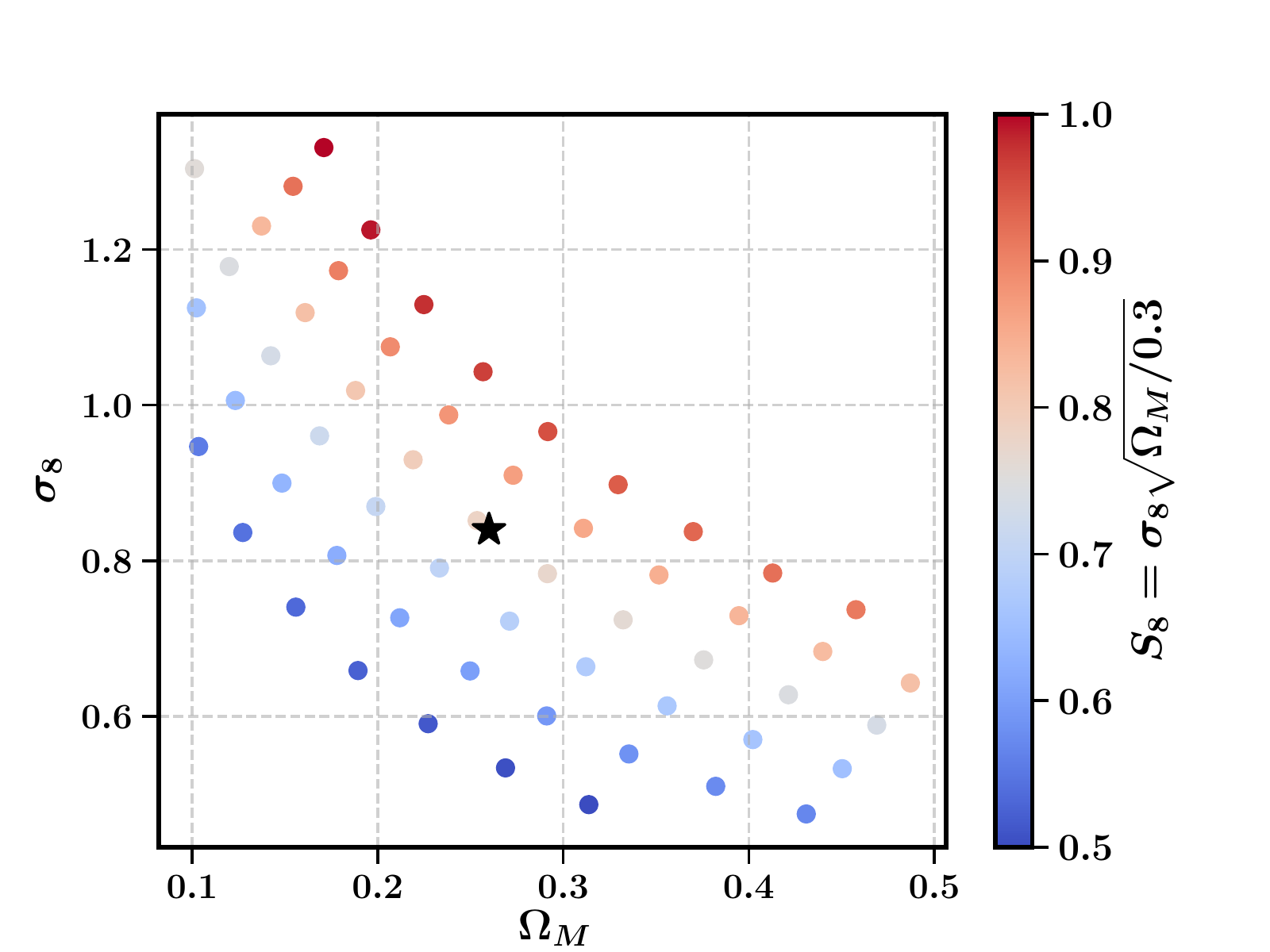}
\caption{\label{fig:simulation_grid} Distribution of the cosmologies in
the $\Omega_{\mathrm{m}} - \sigma_8$ plane sampled with \texttt{PKDGRAV3} dark-matter-only N-Body simulations.
The color corresponds to the value of the $S_8$ parameter at each point and the star
denotes the location of the fiducial cosmology. For the fiducial cosmology 50 simulations
were generated, whereas 5 simulations were ran for all the other cosmologies.}
\end{figure}

\noindent Following \cite{fluri2019cosmological}, we chose to distribute the sampled cosmologies
in the $\Omega_{\mathrm{m}} - \sigma_8$ plane
along lines of approximately constant $S_8$, centered at the DES Y1 cosmic shear results.
We run 50 simulations with different initial conditions for the fiducial cosmology
($\Omega_{\mathrm{m}}=0.26$, $\sigma_8=0.84$) and 5 simulations for each other cosmology.
The simulation grid is shown in Figure \ref{fig:simulation_grid}. \\

\subsection{Mass Map Creation}
\label{sec:mass_maps}
We obtain projected WL mass maps from the N-Body simulations using the
\texttt{UFalcon}\footnote{\url{https://cosmology.ethz.ch/research/software-lab/UFalcon.html}} package.
A detailed description of \texttt{UFalcon} is given in \cite{sgier2019fast}.
We use \texttt{UFalcon} to build a mass map from the discrete particle density shells of the simulations using a method
developed in \cite{teyssier2009full, pires2009cosmological,
schmelzle2017cosmological}. The contribution to the mass map by a single
source at redshift $z_s$ is calculated as
\begin{equation}
\kappa(\hat{n}) = \frac{3 \Omega_{\mathrm{m}}}{2} \int_0^{z_s} \frac{dz}{E(z) a(z)}\frac{\mathcal{D}(0,z) \mathcal{D}(z,z_s)}{\mathcal{D}(0,z_s)}\delta\left(\frac{\mathrm{c}}{\mathrm{H}_0} \mathcal{D}(0,z)\thinspace \hat{n},z\right),
\label{eq:kappa}
\end{equation}
where $\delta$ denotes the density contrast at redshift $z$ projected onto the sphere
along the line-of-sight $\hat{n}$. We introduced the dimensionless
comoving distance $\mathcal{D}(z_1, z_2)$ between two redshifts $z_1$ and $z_2$.
The function $E(z)$ is defined as
\begin{equation}
d\mathcal{D} = \frac{dz}{E(z)}.
\end{equation}
Note that \texttt{UFalcon} avoids a full ray-tracing treatment by utilizing the Born approximation.
By approximating the integral in Equation \ref{eq:kappa} as a discrete sum over shells
of finite thickness in redshift space one can write
\begin{equation}
\kappa(\hat{n}) \approx \frac{3 \Omega_{\mathrm{m}}}{2} \sum_b W_{\mathrm{b}} \int_{\Delta_{z_{\mathrm{b}}}} \frac{dz}{E(z)}\delta\left(\frac{\mathrm{c}}{\mathrm{H}_0} \mathcal{D}(z)\thinspace \hat{n},z\right),
\end{equation}
where the weighted contribution $W_{\mathrm{b}}$ of each shell is given by
\begin{equation}
W_{\mathrm{b}} = \frac{\int_{\Delta_{z_{\mathrm{b}}}} \frac{dz}{E(z) a(z)} \frac{\mathcal{D}(0,z) \mathcal{D}(z,z_{\mathrm{s}})}{\mathcal{D}(0,z_{\mathrm{s}})}} {\left(\int_{\Delta_{z_{\mathrm{b}}}} \frac{dz}{E(z)}\right)}.
\end{equation}
To consider a continuous distribution of sources in redshift space, described by
$n(z)$, the shell weights need to be modified to
\begin{equation}
W_{\mathrm{b}}^{n(z)} = \frac{\int_{\Delta_{z_{\mathrm{b}}}} dz \int_z^{z_{\mathrm{f}}} dz' \frac{n(z')}{a(z) E(z)}\frac{\mathcal{D}(0,z)\mathcal{D}(z,z')}{\mathcal{D}(0,z')}}
{\left(\int_0^{z_{\mathrm{f}}} dz \thinspace n(z)\right) \left(\int_{\Delta_{z_{\mathrm{b}}}} \frac{dz}{E(z)}\right)},
\label{eq:lens_weight}
\end{equation}
where $z_{\mathrm{f}} = 3.0$ in our setup.
By pixelizing the sphere into $N_{\mathrm{pix}}$ pixels of equal area and using
that the density contrast $\delta$ can be related to the pixel-averaged particle density
$n_{\mathrm{p}}(\hat{n}_{\mathrm{pix}},z)$ at redshift $z$,
measured within the pixel in direction $\hat{n}_{\mathrm{pix}}$,
one can write down a pixelized version of the projected density contrast
$\delta\left(\frac{\mathrm{\mathrm{c}}}{\mathrm{H}_0} \mathcal{D}(z)\thinspace \hat{n},z\right)$ as
\begin{equation}
\delta\left(\frac{\mathrm{c}}{\mathrm{H}_0} \mathcal{D}(z)\thinspace \hat{n}_{\mathrm{pix}},z\right) = \left(\frac{\mathrm{H}_0}{c}\right)^3 \frac{V_{\mathrm{sim}} N_{\mathrm{pix}}}{4\pi N_{\mathrm{part}}}\frac{n_{\mathrm{p}}(\hat{n}_{\mathrm{pix}}, z)}{\mathcal{D}^2(0,z)},
\end{equation}
as outlined in \cite{sgier2019fast}. In our work, we adapt $(900 \thinspace \mathrm{Mpc}/h)^3$
for the simulation volume $V_{\mathrm{sim}}$ and $768^3$ for the number of simulated
particles $N_{\mathrm{part}}$. To pixelize the sphere, the \texttt{HEALPIX} package
\cite{gorski1999healpix} is used with a pixel resolution of $\texttt{NSIDE}=1024$
leading to $N_{\mathrm{pix}}=12 \cdot \texttt{NSIDE}^2$.
Please refer to \cite{sgier2019fast} for a more detailed description of this procedure. \\

\begin{figure}
\centering
\includegraphics[width=.6 \textwidth]{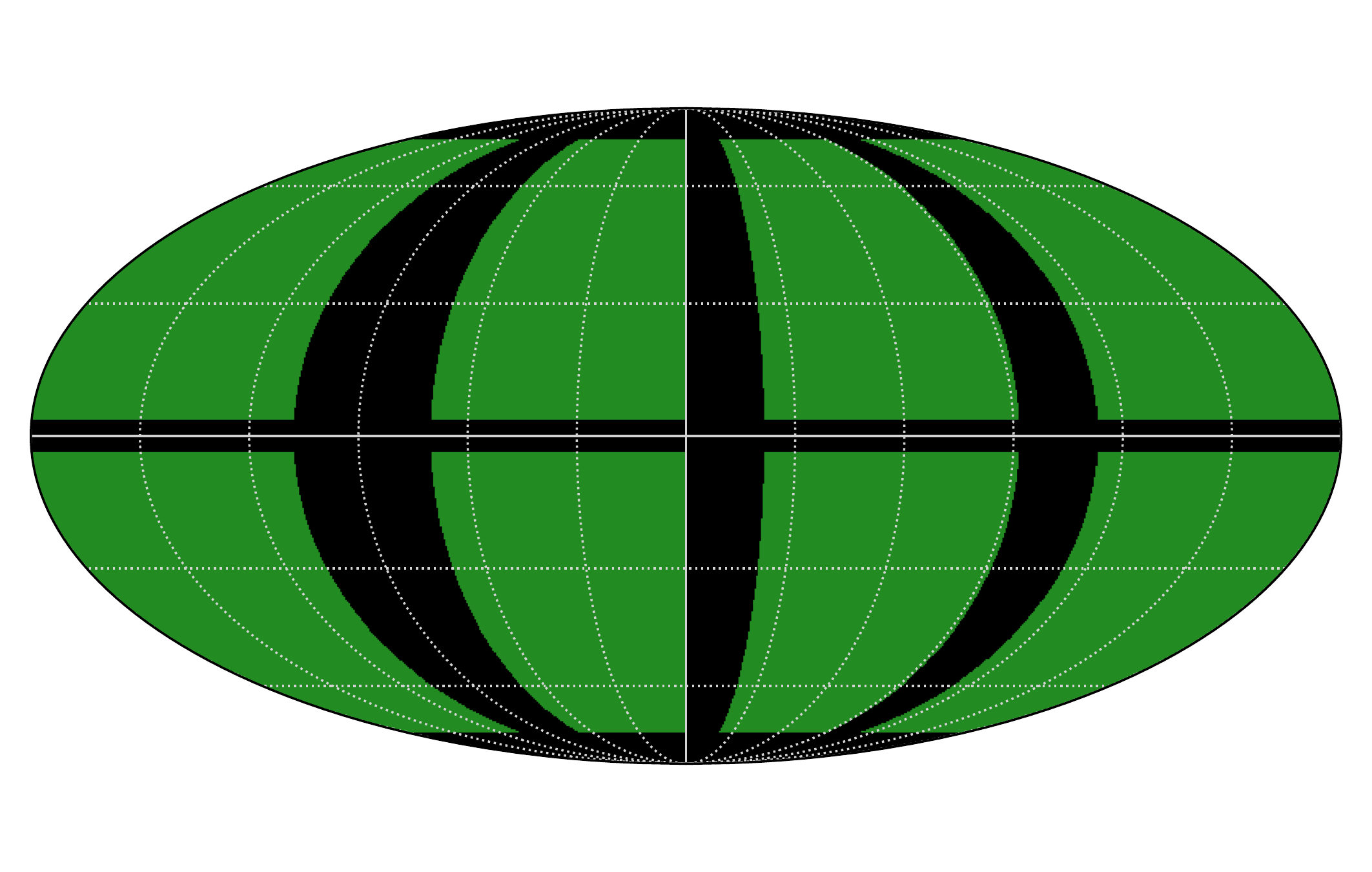}
\caption{\label{fig:mask} The 8 survey masks of the rotated mock surveys.
Each one of the 8 rotated surveys spans an angular area of 5000 deg$^2$.}
\end{figure}

\noindent The mass maps created from the N-Body simulations using \texttt{UFalcon}
span the full sky, containing the cosmological signal only.
To produce mass maps with the same survey properties as the mock, we need
to cut out patches from the simulated full-sky mass maps that have the same
shape and size as the mask of the mock survey.
To do so, we rotate the galaxy positions on the sky, which allows us to produce 8
simulated surveys from a single \texttt{PKDGRAV3} simulation. We checked that the
rotation does not introduce any artifacts, by comparing the angular power spectra
of the masks of the rotated surveys.
The distribution of the 8 survey masks on the sky is shown in Figure \ref{fig:mask}. \\

\noindent At this point, the mass maps only contain the cosmological signal.
We need to add a noise component to optimally reconstruct the survey properties of the mock survey .
To do so we first convert the simulated mass map
$\kappa_{\mathrm{sim}}$ to a shear field $\gamma_{\mathrm{sim}}$, using the spherical
Kaiser-Squires mass-mapping method (see Equation \ref{eq:KS}).
The noise signal is drawn from the mock survey by randomly rotating
the ellipticities of the galaxies in place (according to Equation \ref{eq:rotation})
and added to the cosmological shear signal
$\gamma_{\mathrm{sim}}$ on the pixel level according to
\begin{equation}
\gamma_{\mathrm{pix}} = \gamma_{\mathrm{noise}} + \gamma_{\mathrm{sim}} = \frac{1}{N}\sum_{j=1}^N \gamma_{j,\mathrm{mock}}\exp(i\phi_j) + \gamma_{\mathrm{sim}},
\end{equation}
with the sum running over all $N$ galaxies in the mock survey that are located
in the corresponding pixel. The angles $\phi_j$ are drawn uniformly from the interval
$[0, 2\pi[$. We repeat this procedure 10 times for each \texttt{PKDGRAV3} simulation,
which provides us with $8 \cdot 10 = 80$ realizations per simulation.
Lastly, we convert the shear field back to a mass map, using the spherical
Kaiser-Squires mass-mapping method once again. \\

\subsection{Cosmological Parameter Inference}
\label{sec:inference}
We infer the constraining power in the $\Omega_{\mathrm{m}} - \sigma_8$ plane of the
studied statistics on the basis of a stage-3-like WL survey.
The measurement data-vector $\vec{X}$ is drawn from the simulations at the
fiducial cosmology setting.
As suggested by \cite{fluri2018weak}, we calculate the Figure-of-Merit (FoM)
according to Equation \ref{eq:FoM} for all our fiducial realizations $\vec{X}_{\mathrm{i}}$ of the data-vector and we choose
the realization which yields the FoM closest to the mean of the distribution of
the FoMs as the measurement $\vec{X}$. \\

\noindent We asses the constraining power in the $\Omega_{\mathrm{m}} - \sigma_8$ plane for
the different statistics using Bayesian inference and under the assumption
that the data-vector $\vec{X}$ is drawn from a multivariate
Gaussian distribution, characterized by a mean data-vector $\vec{X}_{\mathrm{M}}$ and a
covariance matrix $\Sigma$. We estimate the covariance matrix $\Sigma$ from the
simulated data-vectors $\vec{X}_{\mathrm{i}}$ at the fiducial cosmology as
\begin{equation}
\hat{\Sigma} = \frac{1}{N_{\mathrm{f}} - 1} \sum_{i=1}^{N_{\mathrm{f}}} (\vec{X}_{\mathrm{i}} - \hat{\vec{X}}_{\mathrm{M}}) (\vec{X}_{\mathrm{i}} - \hat{\vec{X}}_{\mathrm{M}})^T,
\end{equation}
where $\hat{\vec{X}}_{\mathrm{M}}$ denotes the estimate of the mean data-vector $\vec{X}_{\mathrm{M}}$
at the fiducial cosmology and $N_{\mathrm{f}}=4000$ the number of fiducial
realizations. \\

\noindent Since we do not analytically predict the covariance matrix, nor the
data-vectors, but we estimate them from simulations, the likelihood is not
most accurately modelled by a Gaussian likelihood.
As pointed out by \cite{sellentin2015parameter}, the use of an
estimated covariance matrix requires a modification of the likelihood in order
to stay unbiased. The estimation of the data-vectors from a finite number of
simulations instead of using an exact prediction requires a further correction,
that takes into account the additional variance \cite{jeffrey2019parameter}.
Our final likelihood reads
\begin{equation}
L(\vec{X} | \Omega_{\mathrm{m}}, \sigma_8) \propto \left( 1 + \frac{N_{\mathrm{g}}}{(N_{\mathrm{g}} + 1)(N_{\mathrm{f}} - 1)} (\vec{X} - \hat{\vec{X}}_{\Omega_{\mathrm{m}}, \sigma_8})^T\hat{\Sigma}^{-1}(\vec{X} - \hat{\vec{X}}_{\Omega_{\mathrm{m}}, \sigma_8}) \right)^{-N_{\mathrm{f}}/2},
\end{equation}
where $N_{\mathrm{g}}=400$ indicates the number of realizations used to estimate
the data vector $\hat{\vec{X}}_{\Omega_{\mathrm{m}}, \sigma_8}$ at the cosmology in question
and $N_{\mathrm{f}}=4000$ denotes the number of realizations used to
estimate the covariance matrix $\Sigma$ at the fiducial cosmology.

\noindent We use the Markov Chain Monte Carlo (MCMC) ensemble sampler \texttt{emcee}
\cite{foreman2013emcee} to efficiently sample the parameter space.
We use flat priors ranging from 0.1 to 0.5 for $\Omega_{\mathrm{m}}$ and from 0.3 to 1.4 for $\sigma_8$.
We use the \texttt{scipy.interpolate.SmoothBivariateSpline} interpolator \cite{jones2001scipy},
to evaluate the likelihood function at cosmologies that are not on the grid of
simulated cosmologies. Each element of the data-vector is interpolated individually.
We have confirmed, that the interpolator
succeeds in recovering the data-vectors at cosmologies that are not on the sampled
grid with the necessary precision that we require in our analysis (see Section \ref{sec:interpolator}).

\subsection{Data-Vector Compression}
The evaluation of the likelihood requires the inversion of the
covariance matrix. In some of our setups, especially when investigating
combinations of different statistics, the concatenation of the data-vectors for
different scales, tomographic bins and statistics can lead to long
data-vectors and large covariance matrices. The inversion of such large matrices can
be numerically unstable. In addition, we found that many of the elements of the
data vectors are highly correlated. Therefore, we used a
singular value decomposition (SVD) to reduce the dimensionality, along with the
correlations, using the
\texttt{numpy.linalg.svd} routine \cite{walt2011numpy}. The dimensionality of the resulting data-vectors
was not fixed to a pre-decided value, but was chosen for each combination of statistics
individually by keeping as many SVD components as necessary to capture
99.99999\% of the variance of the different realizations of the data-vector. \\

\subsection{Systematics}
\label{sec:systematics}
One of the challenges in cosmic shear studies
are systematic effects. We decided to include the three dominant effects
affecting WL into our analysis, namely; galaxy intrinsic alignment (IA), multiplicative shear bias (m)
and photometric redshift error ($\Delta_z$). In the following,
we describe these systematics and how they were considered in the analysis in more detail.\\

\paragraph{Galaxy intrinsic Alignment}
\label{sec:ia}
One of the main assumptions in WL studies is that the intrinsic ellipticities of
the source galaxies are uncorrelated. It is known that this assumption does not
hold true in real data due to the intrinsic correlation of the ellipticities of the galaxies with
the large-scale structure and with each other. This effect is referred to as
galaxy intrinsic alignment (IA) and can lead to biases in the inferred
values of the cosmological parameters \cite{heavens2000intrinsic}.
IA can be broken down into two different components; intrinsic-intrinsic (II)
and gravitational-intrinsic (GI) alignment. The II component describes the
correlations between galaxy ellipticities and the large-scale structure and the GI
term refers to the correlations between the ellipticities of foreground and
the sheared background galaxies in a particular region of the sky \cite{heavens2000intrinsic}. \\

\noindent The effect of IA cannot be easily modelled with N-Body simulations.
Instead, we use an approach developed by
\cite{hirata2004intrinsic, bridle2007dark, joachimi2011constraints}
based on the non-linear intrinsic alignment model (NLA) to calculate an
IA-signal, that can be treated as an additive
component to the cosmological signal as
\begin{equation}
\kappa_{\mathrm{tot}} = \kappa + A_{\mathrm{IA}}\thinspace \kappa_{A_{\mathrm{IA}}=1}, \\
\end{equation}
with $A_{\mathrm{IA}}$ denoting the galaxy intrinsic alignment amplitude introduced below.
The IA-signal $\kappa_{A_{\mathrm{IA}}}$ can be obtained from the particle shells of the simulations in
a similar fashion as the cosmological convergence signal $\kappa$ itself.
To do so, the same procedure as used in \texttt{UFalcon} is utilized, but the
weights $W_{\mathrm{b}}^{n(z)}$, given in Equation \ref{eq:lens_weight}, are adapted to describe the
IA-signal instead of the lensing signal
\begin{equation}
W_{\mathrm{b}, \mathrm{IA}}^{n(z)}(A_{\mathrm{IA}}) = \frac{2}{3 \Omega_{\mathrm{m}}} \frac{\int_{\Delta_{z_{\mathrm{b}}}} dz F(z, A_{\mathrm{IA}}) \thinspace n(z)}{\left( \int_{\Delta_{z_{\mathrm{b}}}} \frac{dz}{E(z)}\right) \left( \int_0^{z_{\mathrm{f}}} dz \thinspace n(z)\right)},
\end{equation}
where $F(z, A_{\mathrm{IA}})$ is given by
\begin{equation}
F(z, A_{\mathrm{IA}}) = -A_{\mathrm{IA}} C_1 \rho_{\mathrm{crit}} \frac{\Omega_{\mathrm{m}}}{D_+(z)}\left( \frac{1+z}{1+z_0}\right)^{\eta}\left(\frac{\bar{L}}{L_0}\right)^{\beta},
\end{equation}
with $C_1= 5 \cdot 10^{-14} h^{-2} M_{\odot}^{-1} \mathrm{Mpc}^3$ being a
normalization constant, $D_+(z)$ denoting the normalized, linear growth factor
and $\rho_{\mathrm{crit}}$ the critical energy density of the Universe today.
The parameters $\eta$ and $\beta$ allow to model the redshift and luminosity
dependence, while $A_{\mathrm{IA}}$ takes the role of an amplitude describing the overall
strength of the effect. The redshift and luminosity dependence is modelled around
the arbitrary pivot parameters $z_0$ and $L_0$. $\bar{L}$ denotes the average
luminosity of the source galaxy population.
As in \cite{fluri2019cosmological} and \cite{hildebrandt2017kids}, we do not
consider the redshift and luminosity dependence, which corresponds to fixing
$\eta=\beta=0$. We leave $A_{\mathrm{IA}}$ as a free parameter, that we constrain in our analysis,
using a flat prior ranging from -5 to 5 for $A_{\mathrm{IA}}$, as in T18.

\paragraph{Multiplicative shear bias}
Multiplicative shear bias (m) is another systematic effect that is expected to
potentially bias the inferred values of the cosmological parameters.
It can originate from multiple sources in the data reduction process, such as
noise bias (see e.g. \cite{refregier2012noise}), model bias (see e.g. \cite{bernstein2010shape})
or imperfect PSF corrections (see e.g. \cite{paulin2009optimal}).
We incorporate the effect of multiplicative shear bias in our convergence signal
by modifying the overall scale of the fluctuations as
\begin{equation}
\kappa_m = (1 + m)\kappa_{m=0}.
\end{equation}
We keep m as a nuisance parameter in our analysis and infer its value along with cosmology.
We use a normal prior centered at 0.0 with a standard deviation of 0.02.
The scale of the prior was chosen based on T18, assuming an improvement of
$\approx 20\%$ for a stage 3 WL survey. In the tomographic setup, we adapt one
multiplicative shear bias parameter $\mathrm{m}_{\mathrm{i}}$ for each tomographic bin. \\

\paragraph{Photometric redshift error}
Since WL surveys need to target a large number of galaxies, it is not feasible to
determine their redshift spectroscopically but only photometrically.
This can lead to a biased redshift distribution of the source galaxy population.
As shown in previous studies, such as \cite{hildebrandt2020kids+}, errors in the
redshift distribution can bias the inferred values of cosmological parameters.
We take this effect into account by introducing the nuisance parameter $\Delta_z$,
which describes a global shift of the redshift distribution $n(z)$ as
\begin{equation}
n'(z) = n(z - \Delta_z),
\end{equation}
where $n'(z)$ denotes the shifted redshift distribution of the source galaxies.
We infer the value of $\Delta_z$ in our analysis, using a normal prior centered
at 0.0 with a standard deviation of 0.015, which is motivated based on the
priors used in T18 and assuming an $\approx 20\%$ improvement for a stage 3 WL survey.
In the tomographic setup we use one parameter $\Delta_{z,\mathrm{i}}$ for each tomographic bin. \\

\paragraph{Systematics emulator}
Including the nuisance parameters, we require to sample a 5 (or 11 for a
tomographic setup) dimensional parameter space in the MCMC procedure.
In order to make accurate predictions with the interpolator described in
Section \ref{sec:inference}, a sufficiently dense sampling of the parameter space
is needed. This requires us to run simulations for a number of parameter configurations that is exponentially
increasing with the dimension of the sampled hypercube.
We use an emulator approach to reduce the number of required simulations. \\
The idea is to only simulate a sub-sample of the full simulation grid
and use these simulations to fit a parametric model for each element of the data-vector,
emulating the effect of the systematics on the statistics level directly as
\begin{align}
\vec{d}(\Omega_{\mathrm{m}}, \sigma_8, A_{\mathrm{IA}}, m, \Delta_z) &= \sum_i d_i(\Omega_{\mathrm{m}}, \sigma_8, A_{\mathrm{IA}}, m, \Delta_z) \hat{\vec{e}}_i \nonumber \\
&= \sum_i a^i(\Omega_{\mathrm{m}}, \sigma_8, A_{\mathrm{IA}}, m, \Delta_z) \cdot d_i(\Omega_{\mathrm{m}}, \sigma_8) \hat{\vec{e}}_i,
\label{eq:emulator}
\end{align}
where $d_i(\Omega_{\mathrm{m}}, \sigma_8)$ denotes the value of the $i$th element of the
systematics-free data-vector as obtained by using the interpolator described in
Section \ref{sec:inference} and $a^i(\Omega_{\mathrm{m}}, \sigma_8, A_{\mathrm{IA}}, m, \Delta_z)$
the parametric scale factor, which modifies the interpolated data-vector element
to emulate the effect of the systematics. We note, that the separation in Equation \ref{eq:emulator}
is not physically motivated in general. We checked that this approach
does not introduce significant biases in our results (see Section \ref{sec:emulator}).
Following the Occam's razor, we started with a simplistic
model for $a^i(\Omega_{\mathrm{m}}, \sigma_8, A_{\mathrm{IA}}, m, \Delta_z)$ and continuously increased
its complexity until the accuracy of the predictions fulfilled our
requirements, ending up with a model containing 16 parameters that we fit individually
for each element of the data vector. We describe the emulator as well as the tests
that we performed on it, in more detail in Section \ref{sec:emulator}.

\subsection{Codebase}
\label{sec:codebase}
\begin{figure}
\centering
\includegraphics[width=0.95 \textwidth]{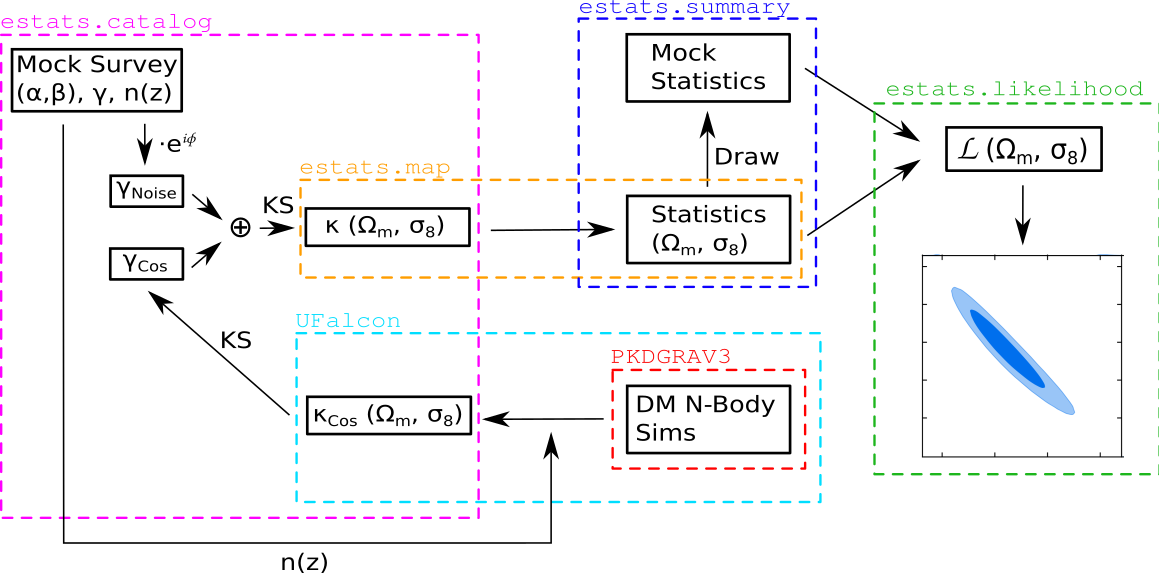}
\caption{\label{fig:flowchart} A schematic illustration of the analysis performed in
this work. We indicate which parts of the code, described in Section \ref{sec:codebase},
are used in which steps of the analysis using the dashed boxes.
The label KS indicates where the Kaiser-Squires map making procedure was used.
The details of the steps are described in detail in Section \ref{sec:methodology}.}
\end{figure}

To allow the reader to reproduce and further understand
the methodology of this work, we publish the repository
\texttt{NGSF}\footnote{\url{https://cosmo-gitlab.phys.ethz.ch/cosmo_public/NGSF}},
containing the pipeline used to run the analysis.
In addition to the main pipeline, we developed three independent packages
as part of this project that are used in the analysis.
We made an effort to organize our codebase in a user-friendly manner, in order to
simplify running such analyses in the future and to make them more accessible and
easily extendable. In particular, we tried to make it easy to implement user-specific,
map-based WL statistics, that can be used in \texttt{estats} and the \texttt{NGSF} pipeline.
The developed packages are \texttt{estats}\footnote{\url{https://pypi.org/project/estats}},
\texttt{esub-epipe}\footnote{\url{https://pypi.org/project/esub-epipe}} and
\texttt{ekit}\footnote{\url{https://pypi.org/project/ekit}}.
All packages are also publicly available on the Python package
index \texttt{PyPi} \footnote{\url{https://pypi.org}}.
The links to the source code and the documentation pages of the packages can be
found in the \texttt{NGSF} repository.
The \texttt{estats} package contains the major building blocks of the pipeline.
The usage of the different \texttt{estats} modules in this analysis is illustrated
in Figure \ref{fig:flowchart}.

\section{Simulated non-tomographic Statistics}
\label{sec:statistics}
We present our findings for the cosmology dependence of the studied non-tomographic statistics;
i.e. we study their dependency on the $S_8$ parameter in Figure \ref{fig:cosmo_curves}.
The colored curves in each panel of Figure \ref{fig:cosmo_curves} show how the
statistics change with $S_8$. The black data points are centered at the
fiducial cosmology, with the error bars indicating the measurement error
expected for a stage-3-like WL survey.
The error bars are estimated from the realizations of the statistics at the fiducial cosmology.
For brevity, we decided to only present the results for one of the
9 considered scales, for the non-Gaussian statistics (see Section \ref{sec:paf_vaf}).
We further present the correlation matrix for the non-tomographic combination
of all studied statistics, namely CLs+PC+MC+MFs, in Figure \ref{fig:corr_mat}.
We refrain from showing the results for the tomographic statistics,
since we believe that their presentation does not provide any additional insights
to the reader. However, we do present the cosmological constraints, that we find
in the non-tomographic, as well as the tomographic setup, in Section \ref{sec:constrains}.
We also summarize and explain the configuration choices, that we made
for the different statistics and their most important features in this section.

\subsection{Angular Power Spectrum (CLs)}
\label{sec:cls}
The angular power spectrum (CLs) is calculated directly from the simulated mass
maps using the \texttt{anafast} routine implemented in \texttt{healpy}
\cite{gorski1999healpix} without performing any prior smoothing of the maps.
Each pixel is weighted by the number of galaxies that fall into its regime.
We consider an angular mutlipole range from $\ell=100$ to $\ell=1000$ using 20
linearly spaced bins. The number of bins was chosen, such that a further division
into more bins does not increase the constraining power any further.
The lower limit of the multipole range ($\ell=100$) was chosen in order
to avoid large-scale regimes, where relativistic corrections become significant
\cite{giblin2017general}. We note that \texttt{PKDGRAV3} actually includes
relativistic corrections and would therefore be suitable to include scales larger
than $\ell=100$, but as one of the goals of this work is to demonstrate that the
addition of non-Gaussian statistics allows to apply conservative scale cuts,
we have decided to not include these scales. To optimally extract the non-Gaussian features
of the maps, we apply a set of smoothing kernels to the maps in order to select
features of different sizes. We found, that when applying a Gaussian smoothing
kernel with a full-width-at-half-maximum (FWHM) of 10.5 arcmin, which corresponds to the smallest scale considered
for the non-Gaussian statistics (see Section \ref{sec:paf_vaf}), the measured CLs drop to $\approx20\%$
compared to the CLs extracted from the unsmoothed maps at a multipole of
$\ell\approx1000$. Hence, we chose $\ell=1000$ as the upper limit for the multipole range,
in order to make the constraints obtained from the CLs more comparable to the ones
from the non-Gaussian statistics. The dependency of the non-tomographic CLs on
$S_8$ is shown in the right, bottom panel of Figure \ref{fig:cosmo_curves}.
In the tomographic setup, we additionally include the cross-power-spectra between
the 4 tomographic bins in the inference process. Note that the CLs measured in this
analysis are not corrected for masking effects and therefore should not be compared
directly to the CLs predicted analytically from a theory code, as these predictions are
typically done for a full-sky map. The parameter constraints in this work are not biased
due to the missing correction for masking effects, since the predictions for the  CLs
at different cosmologies are extracted from the simulations using the same survey mask
as it is used for the mock measurement. The contribution
from the survey noise (indicated by the black, dashed line in Figure \ref{fig:cosmo_curves})
increases on small scales and therefore we obtain the most cosmological constraining power from the
data bins at large scales, below $\ell\approx500$. We can learn from Figure
\ref{fig:corr_mat} that the different data bins of the CLs are highly correlated.

\subsection{Peak and Minimum Counts (PC/MC)}
\label{sec:paf_vaf}
For the non-Gaussian statistics we adapt a multiscale scheme as it was previously
done by \cite{fluri2018weak}. Each mass map is smoothed using 9 different Gaussian
smoothing kernels with a FWHM of
[31.6, 29.0, 26.4, 23.7, 21.1, 18.5, 15.8, 13.2, 10.5] arcmin, respectively.
The statistics are then calculated from each smoothed version of the mass map.
The total data-vector for each statistic is obtained by concatenation of the
9 single scale data-vectors. The application of different smoothing kernels to the
mass maps allows the selection of map features of different sizes.
It was shown, in \cite{fluri2018weak}, that the consideration of even smaller scales
leads to a further increase of the constraining power. However,
on scales smaller than $\approx$ 8 arcmin baryonic effects cannot
be neglected for the PC \cite{weiss2019effects}. Hence, we decided to consider only scales larger than
10 arcmin for all non-Gaussian statistics. \\

\noindent The application of a smoothing kernel to the mass maps washes out the
small-scale fluctuations of the maps and therefore changes the significant range
of $\kappa$ for the PC/MC. Hence, we adapted the range of $\kappa$ used to record
peaks and minima on the mass maps depending on the applied smoothing kernel,
to optimally resolve the distribution of the peaks and minima on all scales.
This complex binning scheme could be avoided by binning the map features in bins
of signal-to-noise ratio (SNR) instead of $\kappa$, as it is was mostly done in
previous studies. However, we found that by doing so the cosmological constraining
power of the PC/MC is diminished. Recording the map features as a function of
SNR instead of $\kappa$, requires to rescale the $\kappa$ values of the extracted features
by the mean standard deviation $<\sigma_{\kappa}>$ of the mass map, estimated on
a pixel level (since the SNR is defined by SNR = $\kappa/<\sigma_{\kappa}> $).
As $<\sigma_{\kappa}>$ itself carries a strong dependency on cosmology,
the PC/MC become more self-similar and cosmological information is lost.
In the case of a combined analysis, using PC/MC and CLs, the lost constraining
power is regained since the cosmology dependence of $<\sigma_{\kappa}>$ is captured
by the CLs. In total we use 15 equally spaced bins per scale.
This number has been chosen, such that increasing the number of bins does not
improve the cosmological constraints anymore.
To suppress the shot noise contribution, we chose the edges of the outermost bins, such that
for each cosmological setup at least 30 peaks/minima are registered in each bin.
The remaining $\kappa$ range is divided into equally spaced bins.
We present the simulated PC/MC for a selected smoothing scale of FWHM=21.1 arcmin
in the top row of Figure \ref{fig:cosmo_curves}. The most information about $S_8$
is obtained from the highest/lowest
$\kappa$ bins of the PC/MC, respectively. The reason being, that such features
of the map are generated by very dense halos and very under-dense
voids, respectively and it is unlikely that such events are produced by random noise.
While the number of less extreme features, as they are recorded in the other data
bins, is certainly larger, the noise contribution (indicated by the black,
dashed line) is dominating in this regime, suppressing the cosmological signal.
With the increase in galaxy density in future surveys, the contribution
from these data bins is expected to grow. From the correlation matrix in
Figure \ref{fig:corr_mat}, we find that
there is some correlation present between the map features found on different
scales for both the PC and MC. This is not unexpected, as the applied smoothing
scales do not differ enough to wash out the features registered on other scales
completely. We find that there is some correlation between the PC/MC and the
CLs, indicating that they partly record similar information of the mass maps.

\subsection{Minkowski Functionals (MFs)}
\begin{figure}
\centering
\includegraphics[width=0.85 \textwidth]{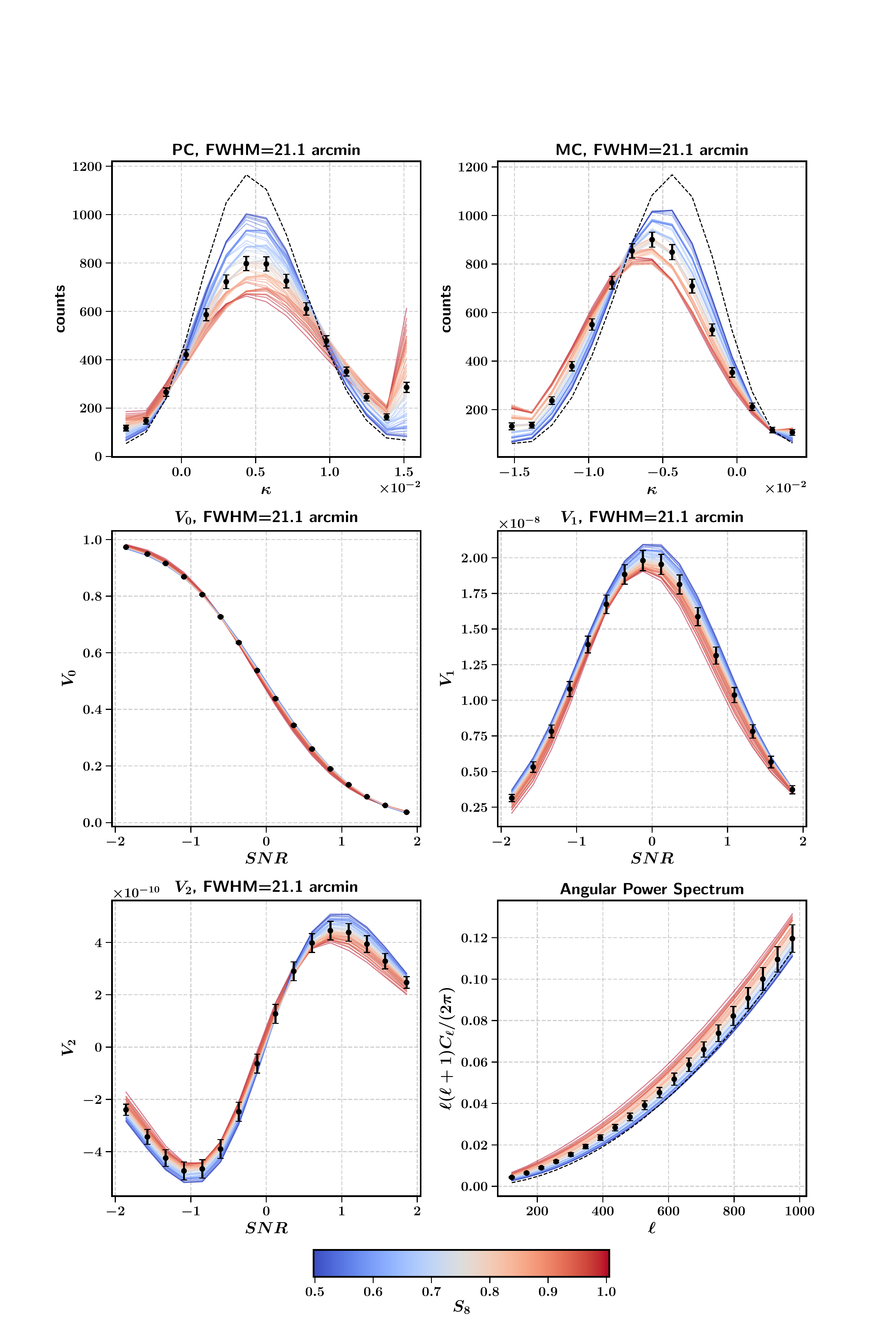}
\caption{\label{fig:cosmo_curves} The simulated statistics recorded in a
non-tomographic setup for varying $S_8$. The two panels at the top illustrate the
$S_8$ dependency of the peak counts (PC) and minimum counts (MC),
binned as a function of the convergence
value $\kappa$ of the detected peaks/minima.
The second row and the left, bottom panel show the simulated
Minkowski functionals (MFs), binned as a function of signal-to-noise ratio (SNR).
For brevity, we decided
to only show the results for one selected scale for the non-Gaussian statistics, namely FWHM=21.1 arcmin.
In the bottom, right panel the angular power spectrum (CLs) is presented.
The black data points shown in each panel are centered at the fiducial cosmology.
The error bars indicate the errors estimated for a stage-3-like WL survey.
For the statistics for which the noise contribution can be understood as an
(approximately) additive component to the cosmological signal, it is
indicated by the black, dashed line.}
\end{figure}

\begin{figure}
\centering
\includegraphics[trim={6cm 0 0 0}, width=0.8 \textwidth, clip]{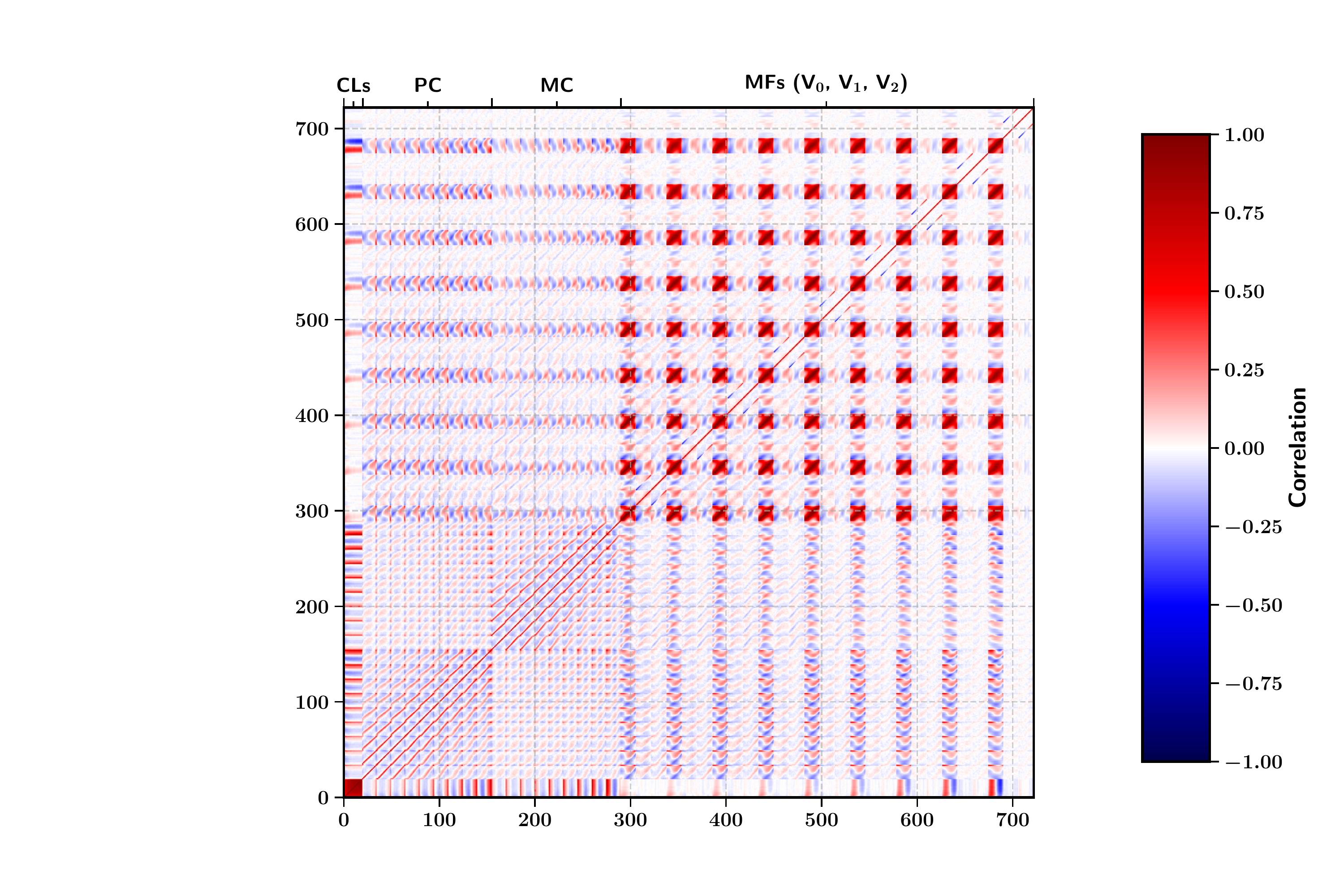}
\caption{\label{fig:corr_mat} The full non-tomographic correlation
matrix at the fiducial cosmology, including all studied statistics,
before applying the singular value decomposition.
The statistics are ordered from left to right as Angular Power Spectrum (CLs),
peak counts (PC), minimum counts (MC) and
Minkowski functionals (MFs) (in order V$_0$, V$_1$ and V$_2$). For the PC, MC and
MFs all 9 scales are concatenated from largest to smallest.}
\end{figure}

\noindent We use 16 different excursion sets $Q_{t}$, spaced linearly in SNR
from -2 to +2 for each scale. We chose the number of considered excursion
sets such that including more does not lead to an increase in the constraining power.
The thresholds $t$ of the sets are chosen in terms of the signal-to-noise ratio (SNR),
i.e. the set $Q_{t}$ contains only pixels with values
$\kappa \geq \nu \cdot <\sigma_{\kappa}>$, where the average standard deviation
of the mass map $<\sigma_{\kappa}>$ is estimated on a pixel level.
As for the PC/MC, we present the MFs for only one selected scale, namely FWHM=21.1 arcmin,
in Figure \ref{fig:cosmo_curves}.
In contrast to the other statistics, we do not show the noise contribution
for the MFs, since the noise cannot be understood as an additive component
on the statistics level, as it is the case for the CLs (and approximately for the PC/MC).
Contrarily to the PC/MC we find less correlation between the MFs and the CLs,
indicating that the MFs probe a different kind of information than the CLs,
PC and MC (see Figure \ref{fig:corr_mat}). \\

\section{Cosmological Constraints}
\label{sec:constrains}
We compare the constraints in the $\Omega_{\mathrm{m}} - \sigma_8$ and $A_{\mathrm{IA}} - S_8$ plane
for the different statistics.
We summarize our findings in Table \ref{tab:parameters} and Table
\ref{tab:parameters_combined}, where we present the constraints on
$\Omega_{\mathrm{m}}$, $\sigma_8$, $S_8$ and $A_{\mathrm{IA}}$ for all statistics and we note the
FoM (Figure-of-Merit) in the $\Omega_{\mathrm{m}} - \sigma_8$ plane, computed as
\begin{equation}
\mathrm{FoM} = \frac{1}{\sqrt{|\hat{\Sigma}(\Omega_{\mathrm{m}}, \sigma_8)|}},
\label{eq:FoM}
\end{equation}
according to T18. The covariance matrix $\hat{\Sigma}$ is estimated from the
MCMC chains. The FoM is anti-proportional to the area of the constraints in the
$\Omega_{\mathrm{m}} - \sigma_8$ plane, with a larger value of the FoM indicating
stronger constraints on $\Omega_{\mathrm{m}}$ and $\sigma_8$. \\

\noindent As a reference, we compare our constraints to the results
found by the Planck 2018 survey \cite{aghanim2018planck} and the DES
cosmic shear analysis from the Year 1 data sample [T18]. We note that a direct
comparison of the constraints found in this work with the results found by T18
is not straightforward, as we have made some different design choices in our analysis.
The main differences include;
1. We use the angular power spectrum, whereas T18 uses 2-point real space correlators,
2. We do not infer the redshift dependence of galaxy intrinsic alignment,
3. We use more conservative scale cuts in our analysis ($\ell \in [100, 1000]$
as opposed to $\theta \in [2.5, 250]$ arcmin in T18).
Given these differences, we suggest the reader to consider the results
found by T18 as a reference for the scale of our constraints only. \\

\noindent All presented constraints were obtained using an MCMC sampling routine
of the parameter space, running 30 chains with an individual length of 50'000 samples. \\

\subsection{Non-tomographic Constraints}
\begin{figure}
\centering
\includegraphics[width=0.8 \textwidth]{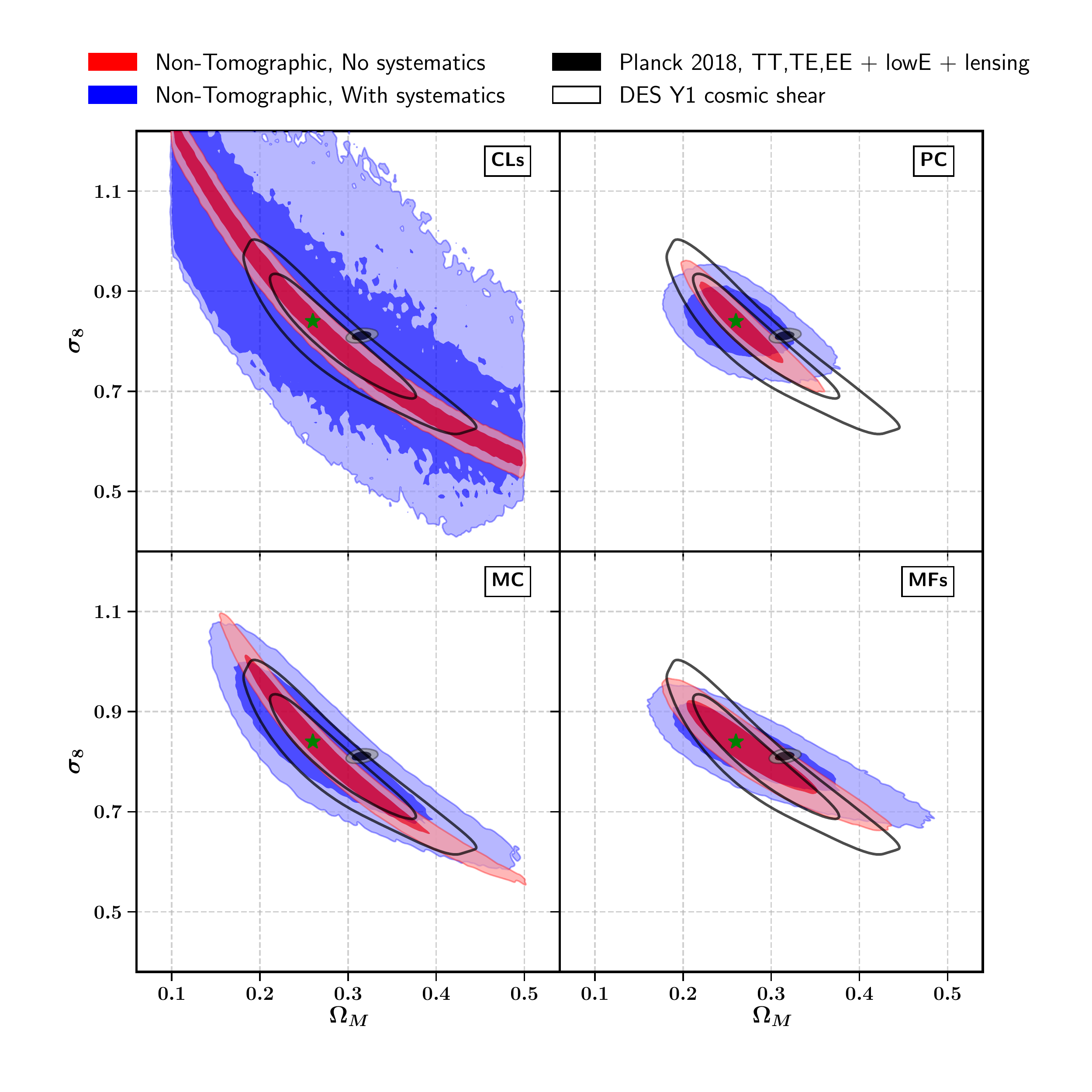}
\caption{\label{fig:contours} Constraints in the
$\Omega_{\mathrm{m}} - \sigma_8$ plane found in the non-tomographic setup. While we have
fixed the nuisance parameters to $A_{\mathrm{IA}}=0, m=0, \Delta_z=0$, when obtaining
the red contours, we find the blue contours when inferring the values of the
nuisance parameters simultaneously with the cosmological parameters.
For reference, we have added the contours found by the Planck 2018
\cite{aghanim2018planck} and DES Y1 cosmic shear \cite{troxel2018dark} surveys.
Note that since we drew the mock measurement in this study from the fiducial
simulations, the contours are centered at the fiducial cosmology, indicated by the green star.
Therefore, the location of our contours should not be compared to the location of the
contours found by DES Y1 and Planck 2018. Only their relative sizes should be compared.
All contours show the 68\% and 95 \% percentiles of the marginalized 2D-distributions.}
\end{figure}

The non-tomographic constraints in the $\Omega_{\mathrm{m}} - \sigma_8$ plane are shown in Figure \ref{fig:contours}.
We recover the typical $\Omega_{\mathrm{m}} - \sigma_8$ degeneracy for the CL
analysis (red contour, upper left panel in Figure \ref{fig:contours}).
For the non-Gaussian statistics, a similar degeneracy is found, although it is
weaker when compared to the CLs.
This also reflects itself in the FoM (see Table \ref{tab:parameters}),
yielding a relative improvement by a factor of $\approx$11, $\approx$3 and $\approx$5
over the CL analysis, for the PC, MC and MFs, respectively.
While for the PC and MC, the direction of the degeneracy is only slightly
different as for the CLs, we record a different degeneracy direction for the MFs,
indicating that the MFs might help to break the degeneracy of the other
statistics in a combined setup (see Section \ref{sec:combined} below).
Overall, we find that all non-Gaussian statistics are less affected by the
$\Omega_{\mathrm{m}} - \sigma_8$ degeneracy and yield stronger constraints than the CL analysis
in a non-tomographic setup and without the consideration of systematic effects. \\

\noindent None of the statistics considered is able to put constraints on
multiplicative shear bias, nor photometric redshift error,
leading to the constraints on $m$ and $\Delta_z$ being prior dominated for all
studied statistics. The uncertainty on $m$ and $\Delta_z$ contributes
to the broadening of the contours in the $S_8$ direction, as seen in
Figure \ref{fig:contours} (blue contours). However, the major contribution to the broadening
in $S_8$ direction originates from the degeneracy with the galaxy intrinsic
alignment amplitude $A_{\mathrm{IA}}$.
All statistics suffer from a loss of constraining
power when systematic effects are included, yielding a deterioration of the FoM
by a factor of $\approx$3, $\approx$3, $\approx$1.8 and $\approx$1.8 for the CLs, PC,
MC and MFs, respectively. We further discuss the non-tomographic constraints on
galaxy intrinsic alignment in Section \ref{sec:IA_contours} below. \\

\subsection{Tomographic Constraints}
\begin{figure}
\centering
\includegraphics[width=0.8 \textwidth]{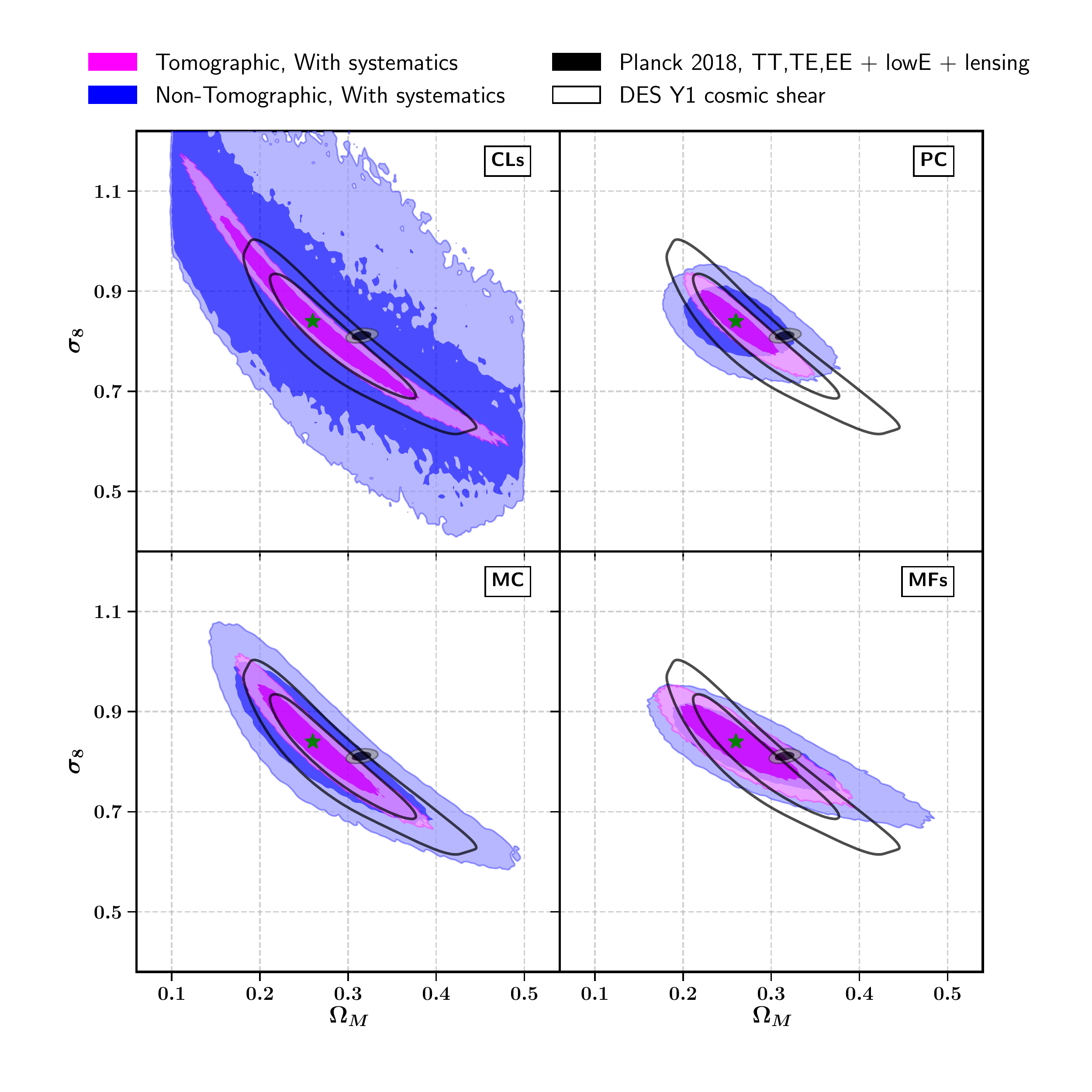}
\caption{\label{fig:contours_tomo} We compare the constraining power of the
different statistics in the $\Omega_{\mathrm{m}} - \sigma_8$ plane when inferring cosmology
and the nuisance parameters simultaneously. While we used a non-tomographic
configuration to obtain the blue constraints, a tomographic setup was used to find
the magenta contours.
For reference, we have added the contours found by the Planck 2018
\cite{aghanim2018planck} and DES Y1 cosmic shear \cite{troxel2018dark} surveys.
Note that since we drew the mock measurement in this study from the fiducial
simulations, the contours are centered at the fiducial cosmology, indicated by the green star. Therefore,
the location of our contours should not be compared to the location of the
contours found by DES Y1 and Planck 2018. Only their relative sizes should be compared.
All contours show the 68\% and 95 \% percentiles of the marginalized 2D-distributions.}
\end{figure}

Since tomography is known to help to improve the robustness of the CL
analysis to galaxy intrinsic alignment, we further study how much the statistics
profit from a tomographic setup. We present the tomographic constraints in Figure
\ref{fig:contours_tomo}.
While we find that tomography increases the constraining power of all studied
statistics, there is a large difference in the improvement between the CLs and the
non-Gaussian statistics, with the non-Gaussian statistics profiting less
(see Table \ref{tab:parameters}). From a comparison of the FoM
between the non-tomographic and tomographic setups (without the consideration of systematics),
we find a relative increase of the FoM, upon introducing tomography, by a
factor of $\approx$4, $\approx$1.2, $\approx$2, $\approx$1.1 for the CLs, PC,
MC and MFs, respectively.
One possible reason for this difference is the fact that the non-Gaussian statistics are
specifically designed to pick up the features of the projected matter field.
Those features become more prominent
as one integrates further over the matter field and sums up the lensing effects
from structures along the line of sight. Therefore, an integration
over a larger redshift range leads to more pronounced over/under-densities on the mass maps.
We also attribute this result to the fact that we include cross-power-spectra
between different tomographic bins in the CL analysis, while we do not consider
such cross-correlations for the non-Gaussian statistics. \\

\noindent While tomography can increase the cosmological constraining power by
providing more information about the three-dimensional structure of the matter field,
its main impact is to constrain galaxy intrinsic alignment, leading to a more pronounced gain in
constraining power over the non-tomographic setup when systematic effects are taken
into account. Again, the CLs profit the most from tomography with a relative increase of
the FoM by a factor of $\approx$10, while the non-Gaussian statistics gain by a
factor of $\approx$3.4, $\approx$3.8 and $\approx$2 for the PC, MC and MFs, respectively
(see magenta contours in Figure \ref{fig:contours_tomo}).
Although the non-Gaussian statistics do not profit from tomography as
much as the CLs do, their cosmological constraining power remains better since they
can extract more cosmological information in the first place.
Note that the non-Gaussian statistics achieve FoM
values without tomography that are similar to those for the CL
analysis with tomography (comparing the blue to the red entries in Table \ref{tab:parameters}).
The PC show the most potential by yielding constraints in the $\Omega_{\mathrm{m}} - \sigma_8$ plane
that are about double in FoM compared to those for the CLs. However, we note
that in the $S_8$ direction the constraints are broader, which is related to
the slightly different direction of the degeneracy and the larger
uncertainty on $A_{\mathrm{IA}}$. \\

\noindent We further discuss the tomographic constraints on
galaxy intrinsic alignment in Section \ref{sec:IA_contours} below. \\

\renewcommand{\arraystretch}{1.5}
\begin{table}
    \centering
    \caption{A comparison of the constraints on all parameters
    (except for $m$ and $\Delta_z$), inferred using the individual statistics.
    The constraints are presented as the median including the 95\% confidence
    intervals in both directions. The fiducial value for each parameter is
    indicated in brackets, on the first line. We also denote the ranges of the
    flat priors that we used in the MCMC procedure for each parameter, on
    the second line. The Figure-of-Merit (FoM) was calculated according to T18.}
    \resizebox{\textwidth}{!}{
    \label{tab:parameters}
    \begin{tabular}{l l l l l l l r l}
    \toprule
     & tomo & sys & $\Omega_{\mathrm{m}}$ (0.26) & $\sigma_8$ (0.84) & $S_8$ (0.79) & FoM & $A_{\mathrm{IA}}$ (0.0) \\
    \midrule
    Prior & - & - & 0.1 -- 0.5 & 0.3 -- 1.4 & - & - &  -5 -- 5 \\
    \midrule
    CLs & N & N & $ 0.29^{+0.20}_{-0.18}$ & $ 0.83^{+0.35}_{-0.28}$ & $ 0.760^{+0.052}_{-0.059}$ & 207 &  -- \\
     & N & Y & $ 0.28^{+0.20}_{-0.18}$ & $ 0.87^{+0.39}_{-0.38}$ & $ 0.80^{+0.33}_{-0.26}$ & 58 & $ 0.2^{+4.1}_{-4.6}$ \\
     & Y & N & $ 0.26^{+0.13}_{-0.12}$ & $ 0.85^{+0.22}_{-0.20}$ & $ 0.776^{+0.027}_{-0.028}$ & 659 &  -- \\
    \vspace{4mm}
     & Y & Y & $ 0.26^{+0.15}_{-0.14}$ & $ 0.85^{+0.23}_{-0.22}$ & $ 0.774^{+0.035}_{-0.035}$ & \textcolor{blue}{485} & $ -0.01^{+0.42}_{-0.44}$ \\
    PC & N & N & $ 0.268^{+0.066}_{-0.060}$ & $ 0.83^{+0.10}_{-0.11}$ & $ 0.781^{+0.030}_{-0.031}$ & 1780 &  -- \\
     &  & Y & $ 0.265^{+0.080}_{-0.075}$ & $ 0.831^{+0.094}_{-0.10}$ & $ 0.777^{+0.096}_{-0.10}$ & \textcolor{red}{586} & $ -0.1^{+1.9}_{-2.2}$ \\
     & Y & N & $ 0.267^{+0.067}_{-0.061}$ & $ 0.833^{+0.087}_{-0.096}$ & $ 0.782^{+0.028}_{-0.029}$ & 2074 &  -- \\
    \vspace{4mm}
     &  & Y & $ 0.265^{+0.060}_{-0.057}$ & $ 0.834^{+0.079}_{-0.089}$ & $ 0.780^{+0.033}_{-0.034}$ & 1964 & $ 0.14^{+0.90}_{-0.84}$ \\
    MC & N & N & $ 0.28^{+0.15}_{-0.12}$ & $ 0.81^{+0.21}_{-0.23}$ & $ 0.774^{+0.041}_{-0.046}$ & 509 &  -- \\
     &  & Y & $ 0.28^{+0.15}_{-0.12}$ & $ 0.82^{+0.20}_{-0.20}$ & $ 0.773^{+0.089}_{-0.094}$ & \textcolor{red}{273} & $ -0.1^{+1.6}_{-2.0}$ \\
     & Y & N & $ 0.268^{+0.099}_{-0.090}$ & $ 0.84^{+0.15}_{-0.14}$ & $ 0.780^{+0.029}_{-0.029}$ & 1097 &  -- \\
    \vspace{4mm}
     &  & Y & $ 0.266^{+0.092}_{-0.084}$ & $ 0.84^{+0.14}_{-0.14}$ & $ 0.779^{+0.036}_{-0.036}$ & 1031 & $ 0.1^{+1.1}_{-1.0}$ \\
    MFs & N & N & $ 0.282^{+0.11}_{-0.094}$ & $ 0.82^{+0.12}_{-0.13}$ & $ 0.785^{+0.049}_{-0.053}$ & 828 &  -- \\
     &  & Y & $ 0.29^{+0.14}_{-0.11}$ & $ 0.82^{+0.11}_{-0.12}$ & $ 0.788^{+0.098}_{-0.10}$ & \textcolor{red}{457} & $ 0.1^{+1.8}_{-1.9}$ \\
     & Y & N & $ 0.271^{+0.092}_{-0.085}$ & $ 0.830^{+0.094}_{-0.10}$ & $ 0.783^{+0.063}_{-0.067}$ & 875 &  -- \\
    \vspace{4mm}
     &  & Y & $ 0.267^{+0.090}_{-0.084}$ & $ 0.832^{+0.095}_{-0.097}$ & $ 0.779^{+0.065}_{-0.070}$ & 895 & $ 0.1^{+1.1}_{-1.0}$ \\
    \midrule
    \makecell{Planck 2018 TT,TE,EE \\ + lowE + lensing} & -- &  -- & $0.315^{+0.015}_{-0.014}$ & $ 0.811^{+0.012}_{-0.012}$ & $ 0.832^{+0.025}_{-0.025}$ & 23170 &  -- \\
    \midrule
    DES Y1, cosmic shear & Y &  Y & $ 0.290^{+0.11}_{-0.094}$ & $ 0.80^{+0.15}_{-0.16}$ & $ 0.778^{+0.055}_{-0.057}$ & 578 & $ 0.8^{+1.3}_{-1.3}$ \\
    \bottomrule
    \end{tabular}}
\end{table}
\renewcommand{\arraystretch}{1}

\subsection{Combined Constraints}
\label{sec:combined}
\begin{figure}
\centering
\includegraphics[width=0.9 \textwidth]{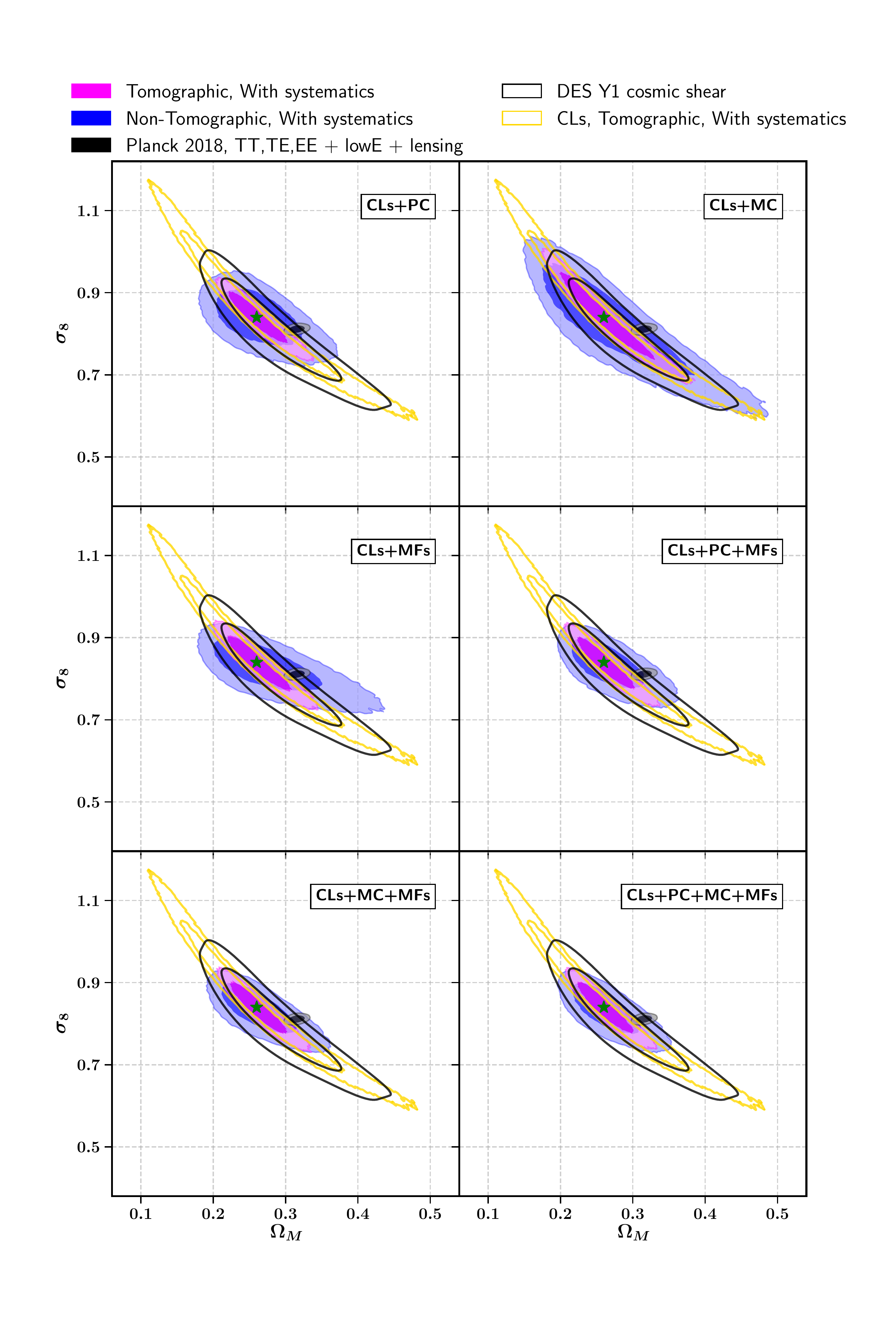}
\caption{\label{fig:contours_combined} We compare the constraining power of
different combinations of statistics in the $\Omega_{\mathrm{m}} - \sigma_8$ plane when
inferring cosmology and nuisance parameters simultaneously.
While we used a non-tomographic configuration to find the blue constraints,
we find the magenta contours using a tomographic setup.
For reference, we have added the contours found by the Planck 2018
\cite{aghanim2018planck} and DES Y1 cosmic shear \cite{troxel2018dark} surveys.
Note that since we drew the mock measurement in this study from the fiducial
simulations, the contours are centered at the fiducial cosmology, indicated by the green star. Therefore,
the location of our contours should not be compared to the location of the
contours found by DES Y1 and Planck 2018. Only their relative sizes should be compared.
All contours show the 68\% and 95 \% percentiles of the marginalized 2D-distributions.}
\end{figure}

With the non-Gaussian statistics probing a different kind
of information of the mass maps than the CLs, we demonstrate that a combination of
the different statistics yields stronger constraints on cosmology, than using the
individual statistics alone.
The constraints obtained when using different combinations of statistics
are presented in Figure \ref{fig:contours_combined}, for both the non-tomographic and
tomographic setups. For comparison, we added the constraints that we find
with the tomographic CL analysis (in yellow). A quantitative comparison of the
constraints is presented in Table \ref{tab:parameters_combined}. \\

\paragraph{Non-tomographic Results} With the CLs and PC carrying the strongest cosmological signal,
we find that combining the two yields tight constraints on the cosmological parameters
in the tomographic and non-tomographic setup (CLs+PC).
However, we observe that
the PC capture nearly all the cosmological information that is recorded by the CLs,
making the contribution of the CLs subdominant in this setup.
The addition of MFs does not increase the constraining power significantly
(CLs+PC+MFs) and neither do the MC (CLs+PC+MC+MFs).
We note however, that a combination of CLs+MC+MFs yields a similar FoM than CLs+PC.
The different direction of the $\Omega_{\mathrm{m}} - \sigma_8$ degeneracy of
the MFs helps to tighten the constraints of the CLs and MC (CLs+MFs and CLs+MC+MFs),
but when the PC are included, yielding constraints that are much smaller than the ones
found with the MFs, the effect becomes negligible (CLs+PC+MC+MFs).
In the described non-tomographic setup, without the consideration of systematic
effects, the addition of the PC / all non-Gaussian statistics to the CLs increases the FoM
by a factor of $\approx$4.1 and $\approx$4.2, respectively. \\

\noindent With the inclusion of systematic effects into the analysis, the contributions
of the MC and MFs become more important, as they are more robust to galaxy intrinsic
alignment. We find that in a non-tomographic setup, the addition of MC and MFs
to the CLs and PC helps to reduce the broadening of the contours
in the $S_8$ direction, that is caused mostly by the uncertainty on $A_{\mathrm{IA}}$
(CLs+PC+MC+MFs). As already noted for the individual
non-Gaussian statistics, we find that a combination of CLs and non-Gaussian statistics
yields tighter constraints in the $\Omega_{\mathrm{m}} - \sigma_8$ plane, without using
tomography, than using the CL analysis with tomography (comparing the
blue to the red entries in Table \ref{tab:parameters_combined}). \\

\paragraph{Tomographic Results} While the non-tomographic combination of all
statistics yields competitive results, the addition of tomography further increases
the cosmological constraining power. When neglecting systematic effects, we
find the same trend as in the non-tomographic case, with the PC
contributing most of the cosmological constraining power, a small contribution by
the CLs and neither the MC nor the MFs contributing significantly for a
combination of all statistics. \\

\noindent When including systematic effects we observe an increase of the FoM
by a factor of $\approx$5.5 as compared to the tomographic CL analysis
and by a factor of $\approx$2.3 over the non-tomographic setup when all statistics
are considered. The gain is mainly achieved by the heightened constraints on $A_{\mathrm{IA}}$,
thanks to the cross-spectra considered in the tomographic CL analysis. With the
tomographic CLs constraining $A_{\mathrm{IA}}$, the role of the MC and MFs in the
tomographic setup is negligible, as their main contribution in the non-tomographic
case was to add robustness to galaxy intrinsic alignment. Again, we leave it to
further studies to investigate if the inclusion of cross-correlations between
different tomographic bins for the non-Gaussian statistics
yields similar constraints on $A_{\mathrm{IA}}$. \\

\renewcommand{\arraystretch}{1.5}
\begin{table}
    \centering
    \caption{Same as Table \ref{tab:parameters}, but comparing the parameter constraints
    using different combinations of the studied statistics.}
    \resizebox{\textwidth}{!}{
    \label{tab:parameters_combined}
    \begin{tabular}{l l l l l l l l l}
    \toprule
     & tomo & sys & $\Omega_{\mathrm{m}}$ (0.26) & $\sigma_8$ (0.84) & $S_8$ (0.79) & FoM & $A_{\mathrm{IA}}$ (0.0) \\
    \midrule
    Prior & - & - & 0.1 -- 0.5 & 0.3 -- 1.4 & - & - &  -5 -- 5 \\
    \midrule
    CLs & Y & Y & $ 0.26^{+0.15}_{-0.14}$ & $ 0.85^{+0.23}_{-0.22}$ & $ 0.774^{+0.035}_{-0.035}$ & \textcolor{blue}{485} & $ -0.01^{+0.42}_{-0.44}$ \\
    \midrule
    CLs + PC & N & N & $ 0.267^{+0.066}_{-0.060}$ & $ 0.83^{+0.10}_{-0.11}$ & $ 0.781^{+0.030}_{-0.031}$ & 1809 & -- \\
     &  & Y & $ 0.265^{+0.078}_{-0.073}$ & $ 0.835^{+0.091}_{-0.096}$ & $ 0.780^{+0.090}_{-0.092}$ & \textcolor{red}{648} & $ 0.0^{+1.8}_{-2.1}$ \\
     & Y & N & $ 0.265^{+0.061}_{-0.057}$ & $ 0.835^{+0.085}_{-0.092}$ & $ 0.781^{+0.023}_{-0.024}$ & 2624 &  -- \\
    \vspace{4mm}
     &  & Y & $ 0.263^{+0.056}_{-0.051}$ & $ 0.836^{+0.077}_{-0.085}$ & $ 0.780^{+0.028}_{-0.028}$ & 2444 & $ 0.02^{+0.36}_{-0.37}$ \\
    CLs + MFs & N & N & $ 0.272^{+0.076}_{-0.070}$ & $ 0.827^{+0.099}_{-0.10}$ & $ 0.782^{+0.045}_{-0.046}$ & 1179 &  -- \\
     &  & Y & $ 0.279^{+0.11}_{-0.092}$ & $ 0.825^{+0.087}_{-0.090}$ & $ 0.789^{+0.10}_{-0.097}$ & \textcolor{red}{597} & $ 0.1^{+1.7}_{-1.9}$ \\
     & Y & N & $ 0.265^{+0.059}_{-0.056}$ & $ 0.834^{+0.086}_{-0.094}$ & $ 0.780^{+0.030}_{-0.031}$ & 2054 &  -- \\
    \vspace{4mm}
     &  & Y & $ 0.264^{+0.059}_{-0.055}$ & $ 0.835^{+0.085}_{-0.088}$ & $ 0.780^{+0.032}_{-0.032}$ & 2040 & $ 0.01^{+0.42}_{-0.45}$ \\
    CLs + MC & N & N & $ 0.28^{+0.13}_{-0.11}$ & $ 0.81^{+0.19}_{-0.21}$ & $ 0.774^{+0.041}_{-0.044}$ & 632 &  -- \\
     &  & Y & $ 0.28^{+0.14}_{-0.12}$ & $ 0.82^{+0.16}_{-0.18}$ & $ 0.772^{+0.083}_{-0.090}$ & \textcolor{red}{337} & $ -0.1^{+1.6}_{-1.9}$ \\
     & Y & N & $ 0.266^{+0.090}_{-0.083}$ & $ 0.84^{+0.14}_{-0.14}$ & $ 0.779^{+0.022}_{-0.023}$ & 1469 &  -- \\
    \vspace{4mm}
     &  & Y & $ 0.264^{+0.087}_{-0.078}$ & $ 0.84^{+0.13}_{-0.13}$ & $ 0.779^{+0.028}_{-0.028}$ & 1363 & $ 0.00^{+0.39}_{-0.41}$ \\
    CLs + PC + MFs & N & N & $ 0.267^{+0.057}_{-0.052}$ & $ 0.830^{+0.093}_{-0.097}$ & $ 0.780^{+0.031}_{-0.032}$ & 2052 &  -- \\
     &  & Y & $ 0.268^{+0.070}_{-0.063}$ & $ 0.831^{+0.076}_{-0.086}$ & $ 0.783^{+0.073}_{-0.073}$ & \textcolor{red}{934} & $ 0.0^{+1.5}_{-1.6}$ \\
     & Y & N & $ 0.264^{+0.056}_{-0.052}$ & $ 0.835^{+0.081}_{-0.088}$ & $ 0.780^{+0.026}_{-0.027}$ & 2524 &  -- \\
    \vspace{4mm}
     &  & Y & $ 0.264^{+0.054}_{-0.049}$ & $ 0.835^{+0.078}_{-0.084}$ & $ 0.780^{+0.028}_{-0.027}$ & 2513 & $ 0.00^{+0.39}_{-0.41}$ \\
    CLs + MC + MFs & N & N & $ 0.268^{+0.058}_{-0.053}$ & $ 0.829^{+0.089}_{-0.10}$ & $ 0.780^{+0.032}_{-0.033}$ & 1973 &  -- \\
     &  & Y & $ 0.265^{+0.071}_{-0.064}$ & $ 0.831^{+0.078}_{-0.082}$ & $ 0.778^{+0.064}_{-0.070}$ & \textcolor{red}{1027} & $ -0.1^{+1.4}_{-1.5}$ \\
     & Y & N & $ 0.265^{+0.056}_{-0.052}$ & $ 0.834^{+0.082}_{-0.088}$ & $ 0.780^{+0.026}_{-0.027}$ & 2487 &  -- \\
    \vspace{4mm}
     &  & Y & $ 0.264^{+0.053}_{-0.050}$ & $ 0.835^{+0.079}_{-0.084}$ & $ 0.780^{+0.028}_{-0.028}$ & 2494 & $ 0.01^{+0.41}_{-0.43}$ \\
    CLs + PC + MC + MFs & N & N & $ 0.267^{+0.056}_{-0.051}$ & $ 0.830^{+0.092}_{-0.096}$ & $ 0.780^{+0.030}_{-0.031}$ & 2199 &  -- \\
     &  & Y & $ 0.265^{+0.064}_{-0.059}$ & $ 0.831^{+0.075}_{-0.086}$ & $ 0.778^{+0.061}_{-0.065}$ & \textcolor{red}{1113} & $ 0.0^{+1.2}_{-1.4}$ \\
     & Y & N & $ 0.264^{+0.054}_{-0.050}$ & $ 0.835^{+0.078}_{-0.084}$ & $ 0.781^{+0.025}_{-0.025}$ & 2695 & -- \\
     &  & Y & $ 0.263^{+0.052}_{-0.049}$ & $ 0.836^{+0.077}_{-0.083}$ & $ 0.780^{+0.028}_{-0.027}$ & 2576 & $ 0.01^{+0.40}_{-0.41}$ \\
    \midrule
    \makecell{Planck 2018 TT,TE,EE \\ + lowE + lensing} & -- &  -- & $0.315^{+0.015}_{-0.014}$ & $ 0.811^{+0.012}_{-0.012}$ & $ 0.832^{+0.025}_{-0.025}$ & 23170 &  -- \\
    \midrule
    DES Y1, cosmic shear & Y &  Y & $ 0.290^{+0.11}_{-0.094}$ & $ 0.80^{+0.15}_{-0.16}$ & $ 0.778^{+0.055}_{-0.057}$ & 578 & $ 0.8^{+1.3}_{-1.3}$ \\
    \bottomrule
    \end{tabular}}
\end{table}
\renewcommand{\arraystretch}{1}

\subsection{Constraints on Galaxy intrinsic Alignment}
\label{sec:IA_contours}
\begin{figure}
\centering
\includegraphics[width=0.9 \textwidth]{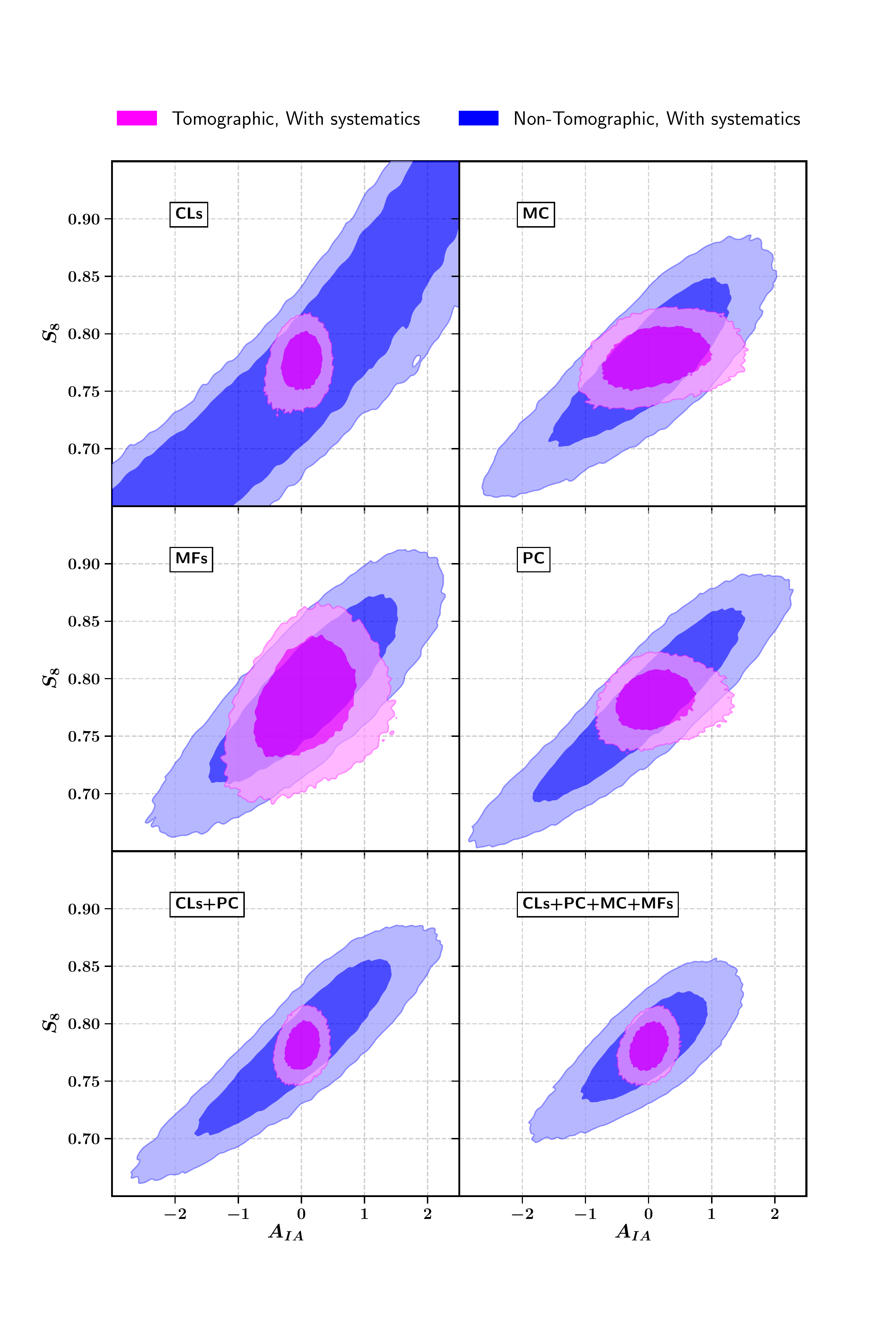}
\caption{\label{fig:contours_IA} We compare the constraining power of
different combinations of statistics in the $A_{\mathrm{IA}} - S_8$ plane when
inferring cosmology and nuisance parameters simultaneously.
While we used a non-tomographic configuration to find the blue constraints,
we find the magenta contours using a tomographic setup.
Note that since we drew the mock measurement in this study from the fiducial
simulations, the contours are centered at the fiducial cosmology.
All contours show the 68\% and 95 \% percentiles of the marginalized 2D-distributions.}
\end{figure}

Figure \ref{fig:contours_IA} shows the constraints in the $A_{\mathrm{IA}} - S_8$ plane for the individual statistics,
as well as for the combination of CLs and MC (CLs+PC) and all statistics (CLs+PC+MC+MFs). \\

\noindent The CLs are unable to constrain $A_{\mathrm{IA}}$ in the
non-tomographic case and their cosmological constraining power is strongly diminished
when $A_{\mathrm{IA}}$ is included in the analysis. This is due to the broadening of the contours in
the $S_8$ direction, caused by the pronounced degeneracy between $S_8$ and $A_{\mathrm{IA}}$ (top, left panel
in Figure \ref{fig:contours_IA}, blue contour).
All non-Gaussian statistics yield better constraints on $A_{\mathrm{IA}}$ in a non-tomographic setup.
However, the PC experience a similar $A_{\mathrm{IA}} - S_8$ degeneracy as the CLs and therefore
a strong broadening of the cosmological constraints along the $S_8$ direction.
On the other hand, we find that the influence of $A_{\mathrm{IA}}$ on the MC and MFs is mildly less
degenerate with $S_8$ and that the cosmological constraints are less affected by the uncertainty
in $A_{\mathrm{IA}}$. One possible explanation for this finding could be, that MFs and especially the MC target
under-dense regions of the matter field, where the effects of galaxy intrinsic
alignment is less dominant due to the lower baryon density. Further, we find that
a combination of different statistics helps to decrease the uncertainty on $A_{\mathrm{IA}}$. \\

\noindent While the CLs are unable to constrain $A_{\mathrm{IA}}$ in a non-tomographic setting,
we find that they yield the strongest constraints on $A_{\mathrm{IA}}$ if tomography is used,
thanks to the consideration of cross-spectra.
We note, that no such cross-correlations
were taken into account for the non-Gaussian statistics.
We leave it to further
studies if the inclusion of such cross-correlations between different tomographic bins
for the non-Gaussian statistics enables them to achieve similar constraints on $A_{\mathrm{IA}}$
as for the CLs.
In this work, we used a simple galaxy intrinsic alignment model to emulate the
effect of galaxy intrinsic alignment on the map level.
We note, that we cannot rule out that our findings regarding galaxy intrinsic
alignment are model dependent and we leave it to further studies to check if
the results change when a more complex galaxy intrinsic alignment model
is used, such as in \cite{blazek2015tidal, blazek2019beyond}.

\section{Conclusions}
\label{sec:conclusions}
We conducted a large-scale simulation
study on the performance of non-Gaussian mass map statistics,
using a realistic stage-3-like WL survey setup.
We compare the constraining power in the $\Omega_{\mathrm{m}} - \sigma_8$ plane
of the angular power spectrum (CLs) with three non-Gaussian statistics,
namely; peak counts (PC), minimum counts (MC) and
Minkowski functionals (MFs). Our analysis features a multiscale scheme to optimally
extract information from the mass maps when using the non-Gaussian statistics.
We compare cosmological constraints in a non-tomographic, as well as a tomographic
setup, using 4 tomographic bins. Furthermore, we investigate on the robustness of the
studied non-Gaussian statistics against the major WL systematic effects, namely;
galaxy intrinsic alignment,
multiplicative shear bias and photometric redshift error. To avoid having to rely
on approximative theory predictions for the non-Gaussian statistics,
that limit their usability, we utilize a forward modelling approach to predict
the statistics based on a suite of dark-matter-only N-Body simulations. \\

\noindent The main findings of this work include:
\begin{itemize}
\item In this setup, we find that the three non-Gaussian statistics considered (PC, MC and MFs) yield stronger
constraints in the $\Omega_{\mathrm{m}} - \sigma_8$ plane when compared to the angular power
spectrum analysis.
They experience a less pronounced $\Omega_{\mathrm{m}} - \sigma_8$ degeneracy.
These findings hold true in a non-tomographic, as well as a tomographic setup.
Taking into account galaxy intrinsic alignment, multiplicative shear bias and
photometric redshift errors does not change this result. In particular, the PC demonstrate
great potential yielding non-tomographic constraints, that are tighter
than the constraints found using tomographic CLs,
even when galaxy intrinsic alignment is taken into account.

\item Including non-Gaussian statistics into the cosmic shear analysis allows us
to apply more conservative scale cuts, while conserving the cosmological constraining power.
This avoids additional uncertainty in the measurement, arising from the influence of
small-scale systematics, in particular baryonic effects.
We find competitive constraints by performing a joined analysis using all four studied
statistics, considering a conservative range of scales ranging from
$\ell=100$ to $\ell=1000$ for the CLs and from 10.5 to 31.6 arcmin
for the non-Gaussian statistics.

\item While the CLs and the PC experience a significant reduction in constraining
power when galaxy intrinsic alignment is taken into account, we find that
the MC and MFs are more resilient to it, thanks to a less pronounced degeneracy between $S_8$ and $A_{\mathrm{IA}}$.

\item We find that the non-Gaussian statistics considered do not profit as
much from a tomographic setup as the CLs do. While the cosmological constraining
power increases considerably for the CLs, the constraints do not tighten up
significantly in case of the non-Gaussian statistics in the absence of systematic effects.
If systematic effects are included all statistics profit from the tomographic setup
due to the improved constraints on galaxy intrinsic alignment.

\item The addition of non-Gaussian statistics to the CLs allows us to find non-tomographic constraints
in the $\Omega_{\mathrm{m}} - \sigma_8$ plane, that are less
than half the size of the constraints found in the tomographic analysis,
taking into account galaxy intrinsic alignment, multiplicative shear bias and
photometric redshift errors.

\item In the context of this study, we developed and distributed a set
of \texttt{Python} software tools aimed at simplifying the production
of such analyses in the future (namely \texttt{NGSF}, \texttt{esub-epipe}, \texttt{estats},
\texttt{ekit}). A short description of the tools is given
in Section \ref{sec:codebase}.
\end{itemize}

\noindent This study explored some alternative WL
statistics in a forward modelling framework.
The introduced simulation framework was developed with a focus
on user-friendliness and extendability, allowing to explore a
multitude of WL statistics, cosmological parameters and systematic effects. \\

\noindent Since we were only able to study the cosmological constraints in the
$\Omega_{\mathrm{m}} - \sigma_8$ plane in this study, we plan to extend the number of investigated
cosmological parameters in the future and to explore which statistics are most
suitable to constrain which parameters. \\

\noindent We plan to extend our study of non-Gaussian statistics, investigating
further statistics such as the profiles around peaks/minima, correlations of peaks/minima
functions or map moments. \\

\noindent So far we have applied conservative scale cuts, mainly in order
to avoid the influence of baryonic effects on small scales.
However, the potential of the information contained
in the non-linear structure of the matter field on small scales is important.
With the non-Gaussian statistics primarily developed to extract this kind of information,
the constraining power could be improved, if these scales were considered.
Therefore, we plan to include a treatment of baryonic effects in
future studies, in order to access the information at smaller scales. \\

\noindent Since none of the non-Gaussian statistics considered was able to put tight
constraints on galaxy intrinsic alignment, except for the CL cross-power-spectra,
we would like to investigate if cross-correlations
between non-Gaussian statistics, measured in different tomographic bins, can put
similar or even tighter constraints on galaxy intrinsic alignment. \\

\noindent Lastly, we note that we have considered a simple galaxy intrinsic
alignment model in this study, neglecting for example the redshift and luminosity
dependence of the effect. We plan to study how the statistics react to a more
complex galaxy intrinsic alignment model.

\acknowledgments
We acknowledge support by grant 200021\_169130 of the Swiss National Science Foundation. \\

\noindent We thank Joachim Stadel and Mischa Knabenhans from University of Z\"urich for
the distribution of \texttt{PKDGRAV3}, as well as their support with the code. \\

\noindent We further thank Jia Liu from University of California, Berkeley, Adam Amara from University of Portsmouth, Aurel Schneider from
University of Z\"urich and the members of the Dark Energy Survey weak lensing working group for the
useful discussions regarding this project. \\

\noindent We would also like to thank Uwe Schmitt from ETH Z\"urich for his support
with the GitLab server and CI engine. \\

\noindent This project used public archival data from the Dark Energy Survey (DES). Funding for the DES Projects has been provided by the U.S. Department of Energy, the U.S. National Science Foundation, the Ministry of Science and Education of Spain, the Science and Technology FacilitiesCouncil of the United Kingdom, the Higher Education Funding Council for England, the National Center for Supercomputing Applications at the University of Illinois at Urbana-Champaign, the Kavli Institute of Cosmological Physics at the University of Chicago, the Center for Cosmology and Astro-Particle Physics at the Ohio State University, the Mitchell Institute for Fundamental Physics and Astronomy at Texas A\&M University, Financiadora de Estudos e Projetos, Funda{\c c}{\~a}o Carlos Chagas Filho de Amparo {\`a} Pesquisa do Estado do Rio de Janeiro, Conselho Nacional de Desenvolvimento Cient{\'i}fico e Tecnol{\'o}gico and the Minist{\'e}rio da Ci{\^e}ncia, Tecnologia e Inova{\c c}{\~a}o, the Deutsche Forschungsgemeinschaft, and the Collaborating Institutions in the Dark Energy Survey. \\
The Collaborating Institutions are Argonne National Laboratory, the University of California at Santa Cruz, the University of Cambridge, Centro de Investigaciones Energ{\'e}ticas, Medioambientales y Tecnol{\'o}gicas-Madrid, the University of Chicago, University College London, the DES-Brazil Consortium, the University of Edinburgh, the Eidgen{\"o}ssische Technische Hochschule (ETH) Z{\"u}rich,  Fermi National Accelerator Laboratory, the University of Illinois at Urbana-Champaign, the Institut de Ci{\`e}ncies de l'Espai (IEEC/CSIC), the Institut de F{\'i}sica d'Altes Energies, Lawrence Berkeley National Laboratory, the Ludwig-Maximilians Universit{\"a}t M{\"u}nchen and the associated Excellence Cluster Universe, the University of Michigan, the National Optical Astronomy Observatory, the University of Nottingham, The Ohio State University, the OzDES Membership Consortium, the University of Pennsylvania, the University of Portsmouth, SLAC National Accelerator Laboratory, Stanford University, the University of Sussex, and Texas A\&M University. \\
Based in part on observations at Cerro Tololo Inter-American Observatory, National Optical Astronomy Observatory, which is operated by the Association of Universities for Research in Astronomy (AURA) under a cooperative agreement with the National Science Foundation. \\

\noindent Based on observations obtained with Planck(http://www.esa.int/Planck),
an ESA science mission with instruments and contributions directly funded by
ESAMember States, NASA, and  Canada. \\

\noindent Some of the results in this paper have been derived using the
\texttt{healpy} and \texttt{HEALPix} packages \cite{gorski1999healpix}. \\

\noindent In this study, we made use of the functionalities provided by
\texttt{numpy} \cite{walt2011numpy}, \texttt{scipy} \cite{virtanen2020scipy}
and \texttt{matplotlib} \cite{hunter2007matplotlib}. \\

\noindent We thank Antony Lewis for the distribution of \texttt{GetDist},
that we relied on to produce some of the plots presented in this work \cite{lewis2019getdist}.

\appendix

\section{Interpolator Test}
\label{sec:interpolator}
\begin{figure}
\centering
\includegraphics[width=1.0 \textwidth]{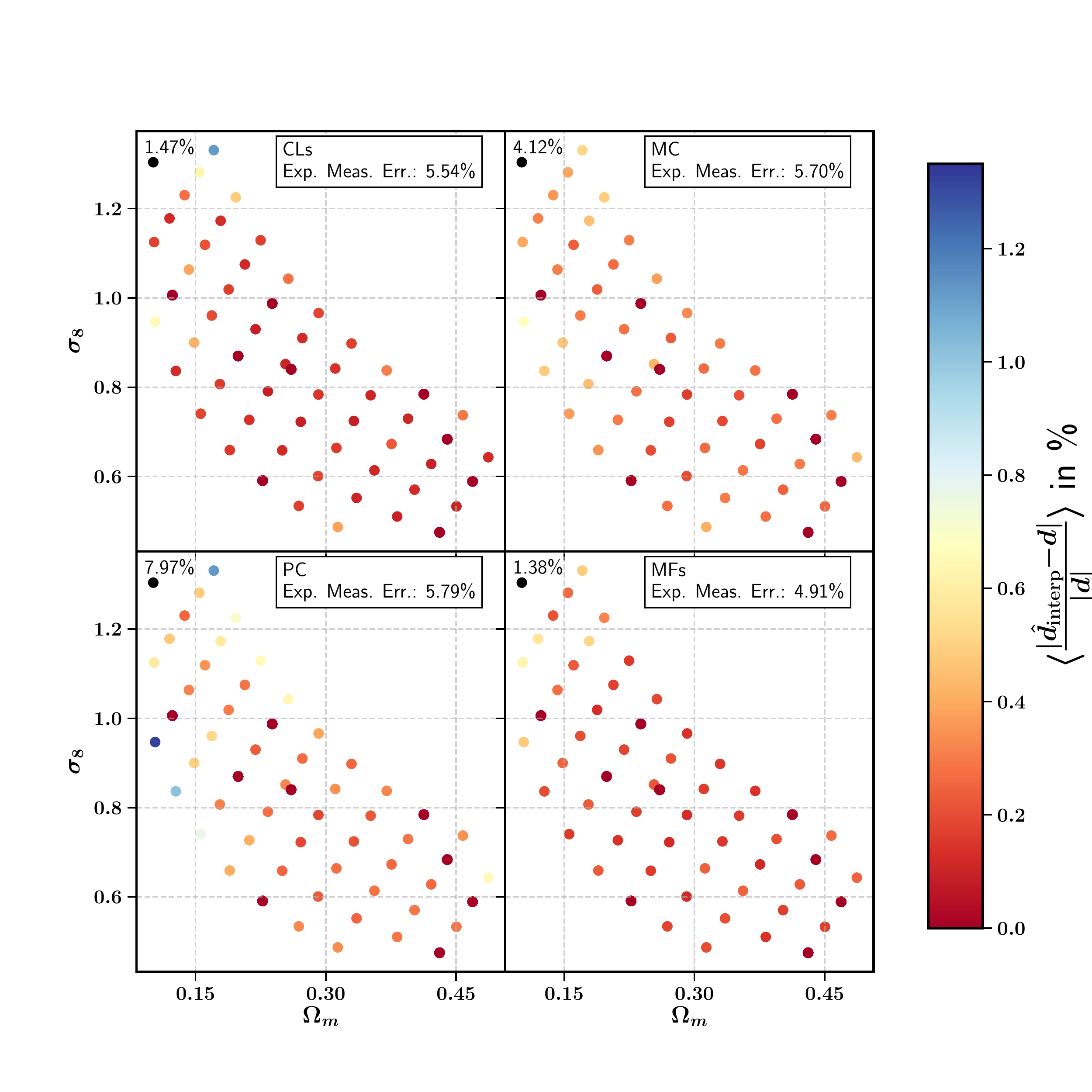}
\caption{\label{fig:interpolator} The results of the test performed
to asses the performance of the interpolator. Each panel shows the grid of cosmologies
sampled with the \texttt{PKDGRAV3} N-Body simulations for a different statistic.
The color indicates the relative error of the interpolator when predicting the
data-vector at this cosmology,
being built on all simulations but the ones at the cosmology in question.
We also denote the estimated measurement error for a stage-3-like survey for each
statistic. Note that the black data point lies outside of the convex hull of
the interpolator and the prediction of the data-vector corresponds to an extrapolation
in this case. The color of this data point does not correspond to the interpolation
error, but is indicated directly.}
\end{figure}

As described in Section \ref{sec:inference}, we use an interpolator to predict
the data-vectors at cosmologies that are not included in the $\Omega_{\mathrm{m}} - \sigma_8$
grid sampled with the \texttt{PKDGRAV3} simulations.
We tested, that the error caused by the interpolator does not bias the
results significantly. The test was performed by building the interpolator using
the simulations for all cosmologies on the sampled grid, except for one cosmology.
We then compared the data-vector $d$ at that remaining configuration,
as predicted by the simulations directly,
to the prediction $d_{\mathrm{interp}}$ of the
interpolator for that missing configuration. We repeated this test for each
cosmology on the simulated grid. The results of this test are visualized in
Figure \ref{fig:interpolator}. We found, that the interpolator succeeds in
recovering the expected data vectors with an error much smaller than the
estimated measurement error for a stage-3-like WL survey for most cosmologies.
Therefore, we conclude that the interpolater is unlikely to bias the results
significantly. The interpolator fails to recover the expected data-vector for one
cosmology only, which is indicated by the black data point in Figure
\ref{fig:interpolator}. We note, that this cosmology is situated outside of the
convex hull of the interpolator, when it is built on the remaining simulations
and therefore it is not expected that the interpolator is able to recover the
data-vector in this specific case.

\section{Emulator Test}
\label{sec:emulator}
In order to overcome the curse of dimensionality and to make our analysis
more easily expandable to a larger grid of cosmological parameters, we develop
a semi-analytical emulator to simulate the effects of the systematics on the
statistic level directly.
In order to avoid biases, caused by the emulator, we require it to recover the
true data-vectors with an error smaller than half of the estimated measurement error for a stage-3-like WL survey. \\

\begin{figure}
\centering
\includegraphics[width=1.0 \textwidth]{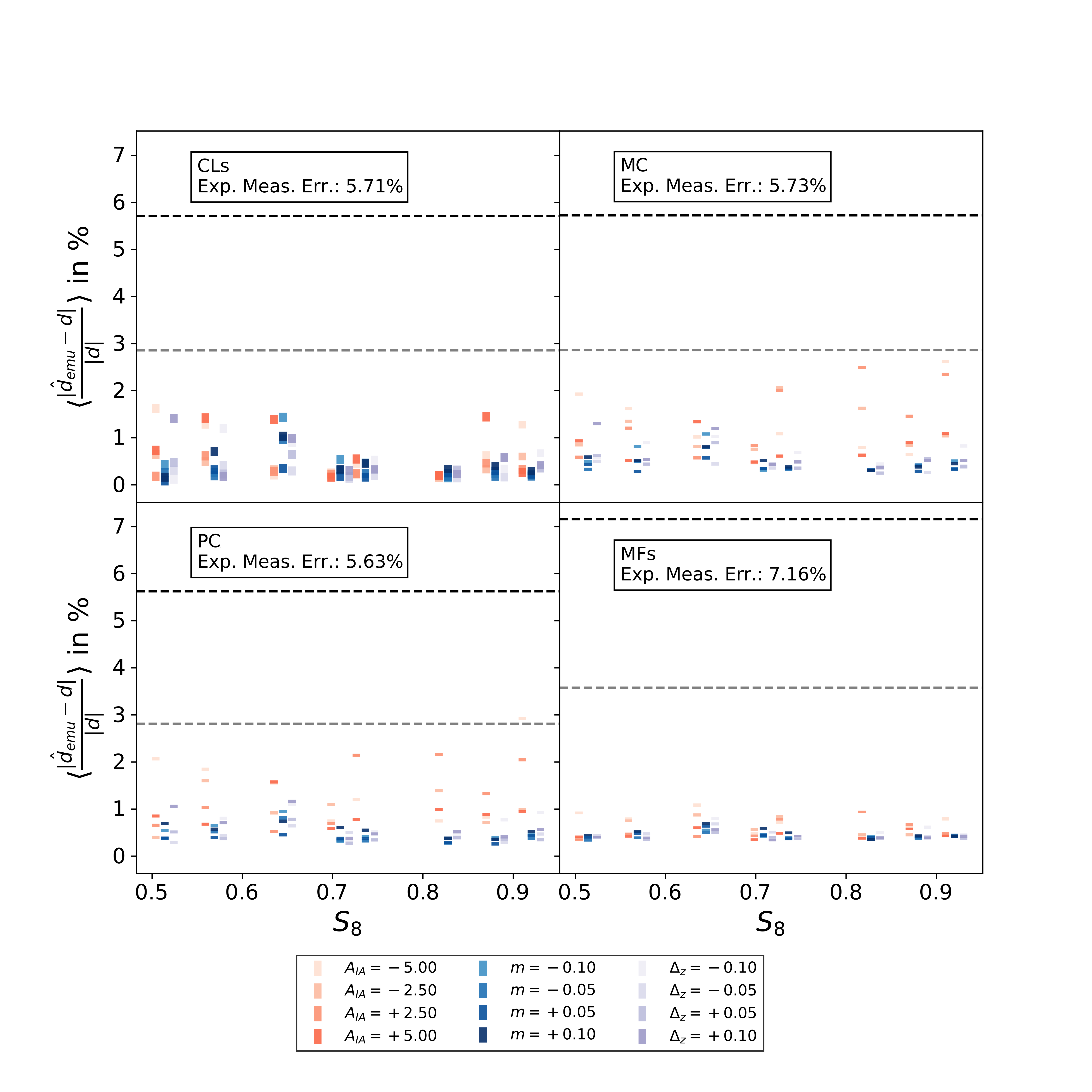}
\caption{\label{fig:emulator_cos} We test how well the emulator is able to
recover the dependency of the different systematics on cosmology. To do so, we
compare the predictions for the mass map statistics for parameter
configurations, that are not included in the samples that are used to build
the emulator to the statistics obtained from the simulations directly.
In each panel, we show the relative difference between the direct simulation and
the emulator prediction. The black dashed line indicates
the estimated measurement error for a stage-3-like WL survey, whereas the grey dashed line indicates
the requirement for the precision of the emulator corresponding to half the
measurement error.}
\end{figure}

\begin{figure}
\centering
\includegraphics[width=1.0 \textwidth]{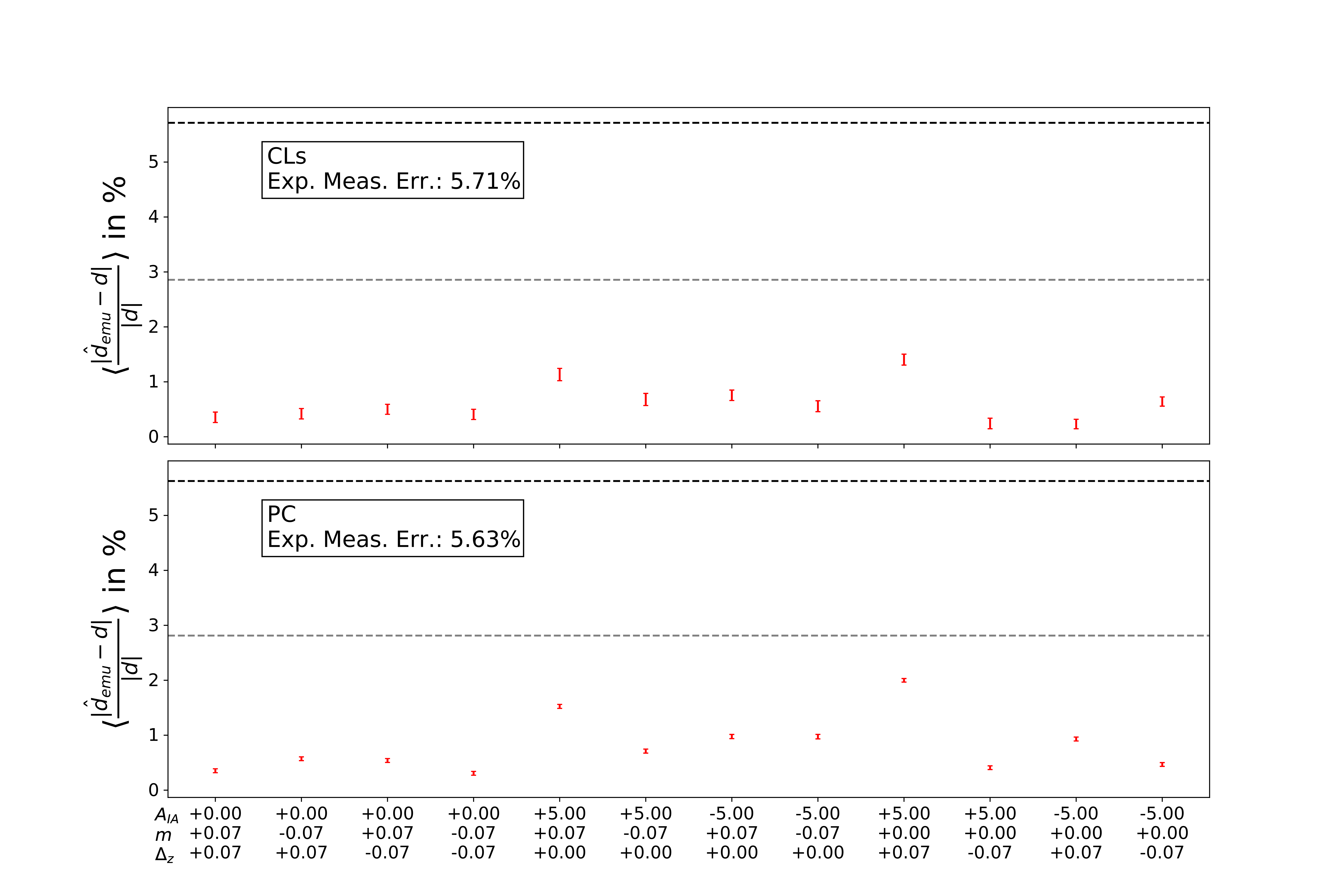}
\caption{\label{fig:emulator_linear_1} The tests assessing how well the
emulator is able to recover the cross-dependencies between different systematics
for the CLs and PC. The test is performed by comparing the predictions for the
mass map statistics for parameter configurations, that are not included in the
samples that are used to build the emulator to the statistics obtained from the
simulations directly. All tested configurations are at the fiducial cosmology.
In each panel we show the relative difference between the direct simulation and
the emulator prediction. The black dashed line indicates
the estimated measurement error for a stage-3-like WL survey, whereas the grey dashed line indicates
the requirement for the precision of the emulator corresponding to half the
measurement error.}
\end{figure}

\begin{figure}
\centering
\includegraphics[width=1.0 \textwidth]{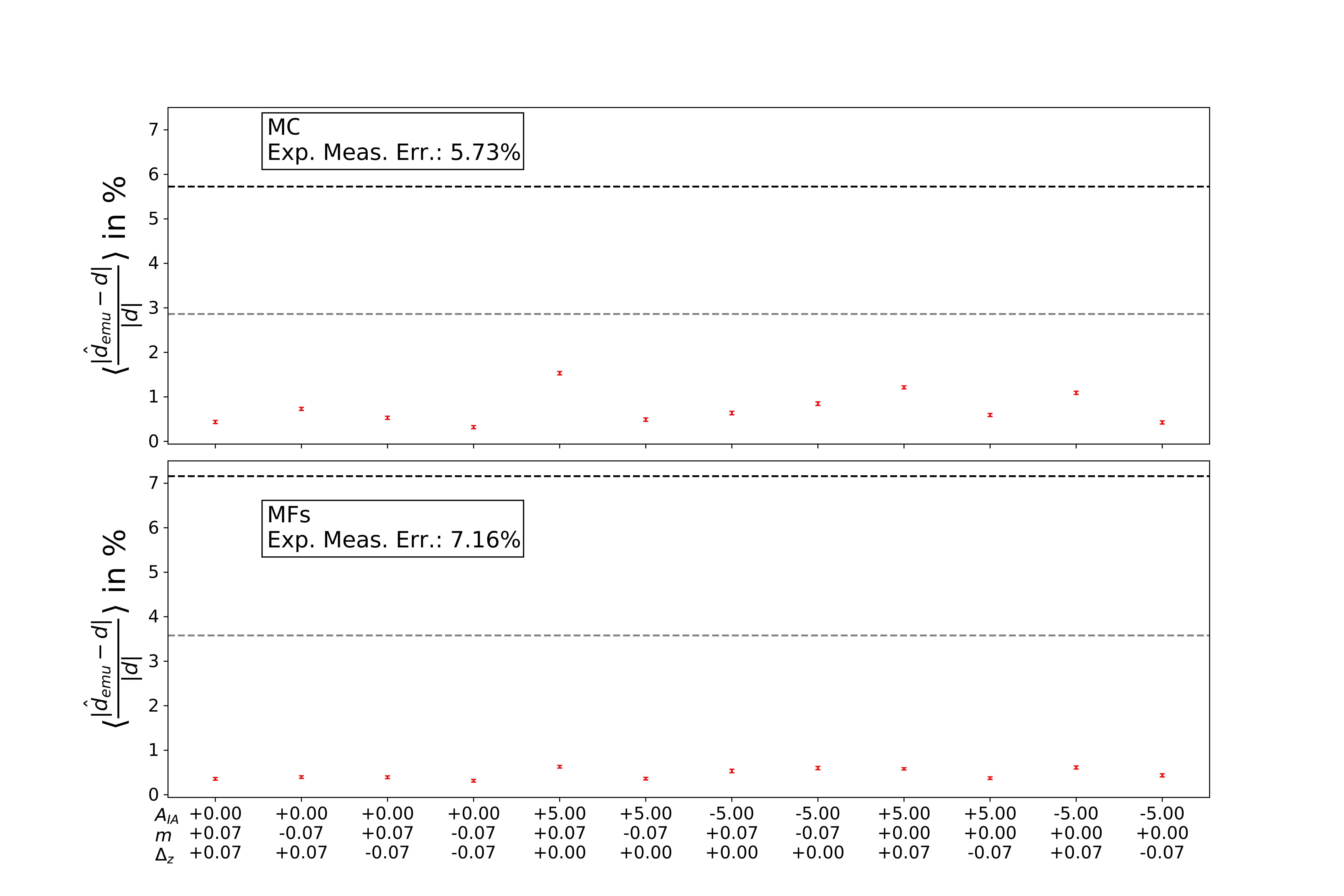}
\caption{\label{fig:emulator_linear_2} Same as Figure \ref{fig:emulator_linear_1},
but for the MC and MFs.}
\end{figure}

\noindent We start from a simple model for the parametric scale factor $a$, introduced
in Equation \ref{eq:emulator}, containing only 3 parameters
\begin{align}
a^i(\Omega_{\mathrm{m}}, \sigma_8, A_{\mathrm{IA}}, m, \Delta_z) &= c^i_1 A_{\mathrm{IA}} + c^i_2 m + c^i_3 \Delta_z,
\end{align}
where the index $i$ denotes an element of the data-vector.
We continuously increased the complexity by adding more terms until the requirement
was met for all statistics, ending up with a model containing 16 parameters
\begin{align}
a^i(\Omega_{\mathrm{m}}, \sigma_8, A_{\mathrm{IA}}, m, \Delta_z) &= c^i_1 A_{\mathrm{IA}} + c^i_2 A_{\mathrm{IA}}^2 + c^i_3 m + c^i_4 m^2 + c^i_5 \Delta_z + c^i_6 \Delta_z^2 \nonumber \\
&+ c^i_7 A_{\mathrm{IA}}^2 \Omega_{\mathrm{m}}  + c^i_8 A_{\mathrm{IA}}^2 \sigma_8 + c^i_9 A_{\mathrm{IA}} \Omega_{\mathrm{m}} \sigma_8 \nonumber \\
&+ c^i_{10} m \Omega_{\mathrm{m}} + c^i_{11} m \sigma_8 + c^i_{12} \Delta_z \Omega_{\mathrm{m}} + c^i_{13} \Delta_z \sigma_8 \nonumber \\
&+ c^i_{14} A_{\mathrm{IA}} m + c^i_{15} m \Delta_z + c^i_{16} A_{\mathrm{IA}} \Delta_z.
\end{align}
We found that a sparse sampling of the nuisance parameter space spanned by the
parameters $A_{\mathrm{IA}}$, $m$ and $\Delta_z$ using $5^3$ configurations suffices to model
the cross-dependency of the nuisance parameters. To model the dependency on
cosmology this sparse sampling of the nuisance parameter space is repeated for
9 different cosmology configurations, distributed on the $\Omega_{\mathrm{m}} - \sigma_8$ grid.
We tested the performance of the emulator by comparing its predictions to the
data-vectors obtained by simulations directly at parameter
configurations that are not included in the sub-sample of configurations that are
used to fit the emulator. The tests yielding the most critical results, probing
the dependency of the systematics on cosmology, as well as the cross-dependency between
the systematics themselves, are shown in Figure \ref{fig:emulator_cos} to
\ref{fig:emulator_linear_2}. The emulator meets our requirements.
However, we note that the modelling of the cosmology dependency and the
dependency on the other systematics for the galaxy intrinsic alignment, for
extreme values of $A_{\mathrm{IA}}$, is not entirely satisfying.
Possible reasons for this effect could be the insufficient complexity of
the emulator for the galaxy intrinsic alignment part or inaccurate
emulation of galaxy intrinsic alignment on the mass map level by the NLA
model for extreme values of $A_{\mathrm{IA}}$.
While the comparisons in this work are not affected significantly by this effect,
it is worth an investigation in future studies.

\bibliographystyle{JHEP.bst}
\bibliography{main.bbl}

\end{document}